\documentclass[12pt]{article}
\usepackage[reqno]{amsmath}
\usepackage{bbm}
\usepackage{epsfig}
\usepackage{array}
\usepackage{float}
\usepackage{dsfont}
\usepackage{amstext}

\usepackage{a4}

\usepackage{a4wide}

\def\ra{\rightarrow}
\def\be{\begin{equation}}
\def\ee{\end{equation}}
\def\gs{\mathrel{
   \rlap{\raise 0.511ex \hbox{$>$}}{\lower 0.511ex \hbox{$\sim$}}}}
\def\ls{\mathrel{
   \rlap{\raise 0.511ex \hbox{$<$}}{\lower 0.511ex \hbox{$\sim$}}}}
\newcommand{\obb}{0\mbox{$\nu\beta\beta$}}
\newcommand{\onbb}{neutrinoless double beta decay }
\newcommand{\ba}{\begin{array}{c}}
\newcommand{\baz}{\begin{array}{cc}}
\newcommand{\bad}{\begin{array}{ccc}}
\newcommand{\bea}{\begin{equation} \begin{array}{c}}
\newcommand{\eea}{ \end{array} \end{equation}}
\newcommand{\ea}{\end{array}}
\newcommand{\D}{\displaystyle}
\newcommand{\dms}{\mbox{$\Delta m^2_{\odot}$}}
\newcommand{\dma}{\mbox{$\Delta m^2_{\rm A}$}}
\newcommand{\meff}{\mbox{$\langle m \rangle$}}
\newcommand{\eV}{\mbox{ eV}}

\newcommand{\sss}{\sin^2 \theta_{12}}
\newcommand{\sch}{\sin^2 \theta_{13}}

\begin{document}

\title{\vspace{-2cm}
\hfill {\small MPP--2006-83}\\
\vspace{-0.3cm} 
\hfill {\small TUM--HEP--638/06}\\
\vspace{-0.3cm} 
\hfill {\small hep--ph/0607103} 
\vskip 0.4cm
\bf 
Low and High Energy Phenomenology of Quark-Lepton Complementarity Scenarios
}
\author{
Kathrin A.~Hochmuth$^a$\thanks{email: \tt hochmuth@mppmu.mpg.de  }~\mbox{ 
}~~and~~
Werner Rodejohann$^b$\thanks{email: \tt werner$\_$rodejohann@ph.tum.de} 
\\\\
{\normalsize \it $^a$Max--Planck--Institut f\"ur Physik
 (Werner--Heisenberg--Institut),}\\
{\normalsize \it F\"ohringer Ring 6, 
D--80805 M\"unchen, Germany}\\ \\ 
{\normalsize \it $^b$Physik--Department, Technische Universit\"at M\"unchen,}\\
{\normalsize \it  James--Franck--Strasse, D--85748 Garching, Germany}
}
\date{}
\maketitle
\thispagestyle{empty}
\vspace{-0.8cm}
\begin{abstract}
\noindent 
We conduct a detailed analysis of the phenomenology of two predictive 
see-saw scenarios leading to Quark-Lepton Complementarity. 
In both cases we discuss the neutrino 
mixing observables and their correlations, neutrinoless double beta decay 
and lepton flavor violating decays such as $\mu \rightarrow e \gamma$. We 
also comment on leptogenesis.  
The first scenario is disfavored on the level of one to two 
standard deviations, in particular due to its prediction for $|U_{e3}|$. 
There can be resonant leptogenesis with 
quasi-degenerate heavy and light neutrinos, which would imply 
sizable cancellations in neutrinoless double beta decay.
The decays $\mu \rightarrow e \gamma$ and $\tau \rightarrow \mu \gamma$ 
are typically observable unless the SUSY masses approach the TeV scale. 
In the second scenario leptogenesis is impossible. It 
is however in perfect agreement with all oscillation data. 
The prediction for $\mu \rightarrow e \gamma$ is in general too large, 
unless the SUSY masses are in the range of several TeV. In this case  
$\tau \rightarrow e \gamma$ and $\tau \rightarrow \mu \gamma$ 
are unobservable.
\end{abstract}

\newpage

\section{\label{sec:intro}Introduction}
The neutrino mass and mixing phenomena \cite{reviews} have provided us with 
some exciting hints towards the structure of the underlying theory of 
flavor. 
In particular, based on observations implying that the 
CKM and PMNS matrices are 
linked by a profound connection, an interesting class of models arises. 
The CKM matrix is to zeroth order the unit matrix plus a small correction, 
given by the sine of the Cabibbo angle, $\sin \theta_C = 0.23$. 
Hence, in the quark sector mixing is absent at zeroth order and 
the deviation from {\it no mixing} is small. 
To make a connection to the lepton sector, it was noted \cite{ichichich} 
that the deviation from {\it maximal mixing} is small. 
Indeed, using the bimaximal \cite{bimax} mixing scenario 
as the zeroth order scheme 
and interpreting the observed deviation from maximal 
solar neutrino mixing as a 
small expansion parameter, one can write \cite{ichichich}: 
\be \label{eq:ich}
|U_{e2}| \equiv \sqrt{\frac 12} \, \left( 1 - \lambda_\nu \right)~.
\ee 
With current experimental information \cite{thomas}, we obtain 
$\lambda_\nu = 0.21^{+0.04, \, 0.08,\, 0.11}_{-0.03, \, 0.07, \, 0.11}$, 
where we have inserted the best-fit values and the 1, 2 and $3\sigma$ 
ranges of the relevant oscillation 
parameters. This number is remarkably similar to 
the Cabibbo angle \cite{ichichich}. In fact, the so called QLC-relation 
(Quark-Lepton Complementarity) \cite{QLC0,QLC1} 
\be \label{eq:QLC}
\theta_{12} + \theta_C = \frac \pi 4 ~
\ee
has been suggested and several situations in which it 
can be realized have been discussed \cite{QLC0,QLC1,QLC2,QLC3,2others}. 
In general, the PMNS matrix is given by $U_\ell^\dagger \, U_\nu$, 
where $U_\nu$ diagonalizes the neutrino mass matrix and $U_\ell$ originates 
from the charged lepton diagonalization. 
Apparently, deviations from maximal $\theta_{12}$ 
as implied by Eqs.~(\ref{eq:ich}, \ref{eq:QLC}) can be 
obtained if the neutrino mass matrix corresponds to bimaximal mixing 
and the charged lepton mass matrix is diagonalized by either the CKM 
or a CKM-like \cite{devbimax,FPR} matrix. 
The opposite case, namely bimaximal mixing from the charged lepton sector 
and a CKM correction from the neutrinos, can also be 
realized, which indicates two possibilities for the approximate 
realization of Eq.~(\ref{eq:QLC}).\\

In the present article we fully analyze the phenomenology 
of these two popular scenarios, proposed in \cite{QLC0,QLC1},  
leading to an approximate realization of QLC within the see-saw 
mechanism \cite{seesaw}. The two scenarios 
show the feature that  the matrix perturbing the bimaximal mixing 
scenario is exactly the CKM matrix and not just a CKM-like matrix, 
which minimizes the number of free parameters. 
We study the neutrino oscillation phenomenology, neutrinoless double 
beta decay and  -- in context of 
the see-saw mechanism -- lepton flavor violating decays such as 
$\mu \to e \gamma$. We present our results of the correlations between 
the observables in several plots. 
In contrast to many previous works, we include the full number of 
possible $CP$ phases. This is a new approach particularly for the 
second scenario, where bimaximal mixing arises from the 
charged lepton sector. For both scenarios we 
comment on the prospects of leptogenesis. 
We begin in Sec.~\ref{sec:form} with an introduction to 
the formalism required to study the observables. In Secs.~\ref{sec:first} 
and \ref{sec:second} we discuss the phenomenology of the two 
scenarios, before we conclude in Sec.~\ref{sec:concl} with a 
summary of our results.

\section{\label{sec:form}Formalism}
In this section we briefly introduce the required formalism to 
analyze the QLC scenarios. 
First, we discuss lepton and quark mixing before 
turning to lepton flavor violation, whose connection to low energy neutrino 
physics is implied by the see-saw mechanism. 
Conclusively, the principles of leptogenesis are outlined.

\subsection{\label{sec:lowNu}Neutrino Masses, Lepton- and Quark-Mixing}
The two scenarios leading to QLC are set within the framework of the 
see-saw mechanism 
for neutrino mass generation \cite{seesaw}. 
In general, one has the Lagrangian  
\be \label{eq:L}
{\cal L} = \frac 12 \, \overline{N_R} \, M_R \, N_R^c + 
\overline{\ell_R} \, m_\ell \, \ell_L + \overline{N_R} \, m_D \, \nu_L~, 
\ee
where $N_R$ are the right-handed Majorana singlets, 
$\ell_{L,R}$ the left- and right-handed charged leptons and 
$\nu_L$ the left-handed neutrinos. The mass matrix of the 
charged leptons is $m_\ell$, $m_D$ is the Dirac 
neutrino mass matrix and $M_R$ the heavy 
right-handed Majorana neutrino mass matrix. 
As $M_R \gg m_D$, Eq.~(\ref{eq:L}) leads to an effective 
neutrino mass matrix at low energies, defined as 
\be \label{eq:mnu}
m_\nu = - m_D^T \, M_R^{-1} \, m_D =  
U_\nu^\ast \, m_\nu^{\rm diag} \, U_\nu^\dagger~,
\ee
where $U_\nu$ transforms $m_\nu$ to $m_\nu^{\rm diag}$, with 
the neutrino masses $m_{1,2,3}$ as diagonal entries. 
When diagonalizing the charged lepton mass matrix as 
$m_\ell = V_\ell \, m_\ell^{\rm diag} \, U_\ell^\dagger$, we can rotate 
$\nu_L \ra U_\nu^\dagger \, \nu_L$, 
$\ell_R \ra V_\ell^\dagger \, \ell_R$ and 
$\ell_L \ra U_\ell^\dagger \, \ell_L$. From the charged 
current term, which is proportional to 
$\overline{\ell_L} \, \gamma^\mu \, \nu_L$, we thus obtain 
the PMNS matrix 
\be \label{eq:pmns_def}
U = U_\ell^\dagger \, U_\nu~,
\ee
which we parameterize as  
\be 
\label{eq:Upara}
U = \left( \bad 
c_{12} c_{13} & s_{12} c_{13} & s_{13} \, e^{-i \delta}  \\[0.2cm] 
-s_{12} c_{23} - c_{12} s_{23} s_{13} e^{i \delta} 
& c_{12} c_{23} - s_{12} s_{23} s_{13} e^{i \delta} 
& s_{23} c_{13}  \\[0.2cm] 
s_{12} s_{23} - c_{12} c_{23} s_{13} e^{i \delta} & 
- c_{12} s_{23} - s_{12} c_{23} s_{13} e^{i \delta} 
& c_{23} c_{13}  \\ 
               \ea   \right) 
 {\rm diag}(1, e^{i \alpha}, e^{i (\beta + \delta)}) \, , 
\ee 
where we have used the usual notations $c_{ij} = \cos\theta_{ij}$, 
$s_{ij} = \sin\theta_{ij}$. We have also introduced the Dirac $CP$-violating
phase $\delta$ and the two Majorana $CP$-violating
phases $\alpha$ and $\beta$ \cite{BHP80}. 
The oscillation parameters can be expressed by two independent 
mass squared differences, $\dms = m_2^2 - m_1^2$ and 
$\dma = |m_3^2 - m_1^2|$, as well as three mixing angles, 
whose exact values are a matter of intense research projects \cite{reviews}. 
Their current best-fit values and their 1, 2 and 3$\sigma$ 
ranges are according to Ref.~\cite{thomas}:  
\begin{eqnarray} \label{eq:data}
\dms &=& 
\left(7.9^{+0.3, \, 0.6, \, 1.0}_{-0.3, \, 0.6,  \,0.8}\right) 
\cdot 10^{-5} \eV^2~,\nonumber\\
\sss &=& 0.31^{+0.02, \,0.06, \, 0.09}_{-0.03, \, 0.05,\, \,0.07} ~,\nonumber\\
\dma &=&  
\left(2.2^{+0.37, \, 0.7, \,1.1}_{-0.27, \, 0.5, \,0.8}\right) 
\cdot 10^{-3} \eV^2~,\\
\sin^2\theta_{23} &=& 0.50^{+0.06, \, 0.14,  \,0.18}
_{-0.05, \, 0.12,  \,0.16} ~,\nonumber\\
\sch &<& 0.012~(0.028, \, 0.046)~.\nonumber
\end{eqnarray}
The present best-fit value for $\sch$ is 0 and there is no information 
on any of the phases.\

Turning to the quark sector, the CKM matrix is \cite{wolf}
\be \label{eq:CKM}
V = \left( 
\bad 
1 - \frac{1}{2} \, \lambda^2 & \lambda & A \, \lambda^3 \, 
(\rho - i \eta) \\[0.3cm]
-\lambda & 1 - \frac{1}{2} \, \lambda^2 & A \, \lambda^2 \\[0.3cm]
A \, \lambda^3 \, (1 - \rho + i \eta) & -A \, \lambda^2 & 1 
\ea
\right) + {\cal{O}}(\lambda^4)~.
\ee
In analogy to the PMNS matrix it is a product of two unitary 
matrices, $V = V_{\rm up}^\dagger \, V_{\rm down}$, where 
$V_{\rm up}$ ($V_{\rm down}$) is associated with the 
diagonalization of the up-(down-)quark mass matrix. 
As reported in \cite{ckm} the best-fit values as well as the 
1, 2 and 3$\sigma$ ranges of the parameters 
$\lambda, A,\bar{\rho},\bar{\eta}$ are  
\bea \label{eq:ckm}
\lambda = \sin \theta_C =  
0.2272^{+0.0010, \, 0.0020, \, 0.0030}
_{-0.0010, \, 0.0020, \, 0.0030}~,\\[0.2cm]
A = 0.809^{+0.014, \, 0.029, \, 0.044}
_{-0.014, \, 0.028, \, 0.042}~,\\[0.2cm] 
\bar\rho = 0.197^{+0.026, \, 0.050, \, 0.074}
_{-0.030, \, 0.087, \, 0.133}~,\\[0.2cm] 
\bar\eta = 0.339^{+0.019, \, 0.047, \, 0.075}_{-0.018, \, 0.037, \, 0.057}~, 
\eea  
where 
$\bar\rho = \rho \,(1 - \lambda^2/2)$ and 
$\bar\eta = \eta \,(1 - \lambda^2/2)$. 
Effects caused by $CP$ violation are always proportional to a 
Jarlskog invariant \cite{JCP}, defined as 
\be \label{eq:jcp}
J_{CP} = -{\rm Im} \{ V_{ud} \, V_{cs} 
\, V_{us}^\ast \, V_{cd}^\ast \} 
\simeq A^2 \, \lambda^6 \, \bar\eta = 
\left( 
3.1^{+0.43, \, 0.82, \, 1.08}_{-0.37 , \, 0.74, \, 0.96}
\right) \cdot 10^{-5}
~.
\ee
The leptonic analogue of Eq.~(\ref{eq:jcp}) is 
\be \label{eq:JCPlep} 
J_{CP}^{\rm lep} = {\rm Im} 
\left\{ U_{e1} \, U_{\mu 2} \, U_{e 2}^\ast \, U_{\mu 1}^\ast \right\} 
= \frac 18 \, \sin 2 \theta_{12} \, \sin 2 \theta_{23} \, 
\sin 2 \theta_{13} \, \cos \theta_{13} \, \sin \delta~,  
\ee
where we have also given the explicit form of $J_{CP}^{\rm lep}$ with the 
parameterization of Eq.~(\ref{eq:Upara}).
There are two additional invariants, $S_1$ and $S_2$ \cite{JMaj87}, 
related to the Majorana phases: 
\bea \label{eq:S1S2}
S_1 = {\rm Im}\left\{ U_{e1} \, U_{e3}^\ast \right\}  
\mbox{ and } 
S_2 = {\rm Im}\left\{ U_{e2} \, U_{e3}^\ast \right\} ~,
\eea
which have no analogue in the quark sector. 

\subsection{\label{sec:LFV0}Lepton Flavor Violation} 
The see-saw mechanism explains the smallness of neutrino masses, but 
due to the extreme heaviness of the right-handed Majorana 
neutrinos a direct test is not only challenging, but presumably 
impossible. Nonetheless a reconstruction of the see-saw parameter 
space is possible in supersymmetric (SUSY) scenarios. 
While being extremely suppressed when mediated by light neutrinos 
\cite{serguey}, 
Lepton Flavor Violating (LFV) decays such as $\mu \ra e \gamma$ 
depend in the context of SUSY see-saw 
on the very same parameters responsible for neutrino masses and can 
be observable in this case \cite{LFV}. 
The size and relative magnitudes of the decays are known to be a useful 
tool to distinguish between different models. 
In this work we will focus on 
models where SUSY is broken by gravity mediation, so called mSUGRA models. 
In this case there are four relevant parameters, which are 
defined at the GUT scale $M_X$, namely the universal scalar mass $m_0$, 
the universal gaugino mass $m_{1/2}$, 
the universal trilinear coupling parameter $A_0$ and $\tan \beta$, which is 
the ratio of the vacuum expectation values of the up- and down-like 
Higgs doublets. 
For the branching ratios 
of the decays $\mu \ra e \gamma$, $\tau \ra e \gamma$ and 
$\tau \ra \mu  \gamma$ one can obtain in the 
leading-log approximation \cite{LFV}
\begin{equation}
\text{BR}(l_i\to l_j   \gamma) \simeq 
\frac{\Gamma(l_i\to e\nu\bar{\nu})}{\Gamma_{\text{total}}(l_i)}
\frac{\alpha_{\text{em}}^3}{G_F^2 \, m_S^8 \, v_u^4}
\left(\frac{3 m_0^2 + A_0^2 }{8\pi^2}\right)^2
\left| \left(\tilde{m}_D^\dagger \, L \, \tilde{m}_D \right)_{ij}
\right|^2 \, \tan^2\beta\;.
\label{eq_ijg}
\end{equation}
Here $v_u = v \, \sin \beta$ with $v = 174$ GeV, 
$m_S$ represents a SUSY particle mass and 
$L = \delta_{ij} \ln M_X/M_i $, with $M_i$ the heavy Majorana masses and 
$M_X = 2 \cdot 10^{16}$ GeV. 
Note that the formulae relevant for lepton flavor 
violation and leptogenesis have to be evaluated in the basis  
in which the charged 
leptons and the heavy 
Majorana neutrinos are real and diagonal. In this very basis we have to 
replace 
\be \label{eq:md_bas}
m_D \rightarrow \tilde{m}_D = V_R^T \, m_D \, U_\ell~,   
\ee  
where $V_R$ diagonalizes the heavy Majorana mass matrix via 
$M_R = V_R^\ast \, M_R^{\rm diag} \, V_R^\dagger$.
The current limit on the 
branching ratio of $\mu \to e \gamma$ is 
$1.2 \cdot 10^{-11}$ at 90\% C.L. \cite{mega}. A future improvement of two 
orders of magnitude is expected \cite{psi}.
In most parts of the relevant soft 
SUSY breaking parameter space, the expression
\begin{align}
m_S^8\simeq 0.5~m_0^2~m_{1/2}^2~(m_0^2 + 0.6 ~m_{1/2}^2)^2\;,
\label{eq_ms}
\end{align}
is an excellent approximation to the results obtained in a full
renormalization group analysis \cite{PPTY03}. 
In order to simplify comparisons of different scenarios, it can be 
convenient to use ``benchmark values'' of the SUSY 
parameters. We choose both pints and slopes of the 
SPS values \cite{SPS} displayed in Table \ref{tab:SPS}.\\

\begin{table}
\begin{center}
\begin{tabular}{|c|c|c|c|c|} \hline 
Point & $m_0$ & $m_{1/2}$ & $A_0$ & $\tan \beta$ \\ \hline \hline 
1a    & 100   & 250       & $-100$  & 10 \\ \hline 
1b    & 200   & 400       & 0       & 30 \\ \hline 
2     & 1450  & 300       & 0       & 10 \\ \hline 
3     & 90    & 400       & 0       & 10 \\ \hline 
4     & 400   & 300       & 0       & 50 \\ \hline 
\end{tabular}
\end{center}
\caption{\label{tab:SPS}SPS Benchmark values for the mSUGRA parameters 
according to Ref.~\cite{SPS}. The values of $m_0$, 
$m_{1/2}$ and $A_0$ are in GeV. 
The slope for Point 1a (2, 3) is 
$m_0 = -A_0 = 0.4 \, m_{1/2}$ ($m_0 = 2 \, m_{1/2} + 850$ GeV, 
$m_0 = \frac 14 \, m_{1/2} - 10$ GeV) with varying  $m_{1/2}$. 
}
\end{table}

In this context it might be worth commenting on renormalization aspects 
of the QLC relation (see also \cite{QLC1}). The running of the 
CKM parameters can always be neglected. However, the case of a 
large $\tan \beta \gs 10$ in the MSSM can imprint sizeable 
effects on the neutrino observables, if the neutrino masses are 
not normally ordered. In our analysis, this would affect only 
the SPS point 4, when the neutrinos have an inverted 
hierarchy or are quasi-degenerate.\\

It proves useful to consider
also the ``double'' ratios,
\bea
  R (21/31) \D
\equiv \frac{\text{BR}(\mu \to e + \gamma)}
  {\text{BR}(\tau \to e + \gamma)}
\simeq 
\frac{\left| (\tilde{m}_D^\dagger \, L \, \tilde{m}_D )_{21}
\right|^2}{\left| (\tilde{m}_D^\dagger \, L \, \tilde{m}_D )_{31}
\right|^2}\;,\\[0.3cm]
 R (21/32) \D 
 \equiv
  \frac{\text{BR}(\mu \to e + \gamma)} {\text{BR}(\tau \to \mu +
    \gamma)}
\simeq 
\frac{\left| (\tilde{m}_D^\dagger \, L \, \tilde{m}_D )_{21}
\right|^2}{\left| (\tilde{m}_D^\dagger \, L \, \tilde{m}_D )_{32}
\right|^2}\;,
\label{DoubleR}
\eea
which are essentially independent of the SUSY parameters.

\subsection{\label{sec:YB0}Leptogenesis} 
Since we will also comment on the possibility of leptogenesis in the 
QLC scenarios, we will summarize the key principles of this mechanism.  
An important challenge in modern cosmology is the explanation of the 
baryon asymmetry $\eta_B \simeq 6 \cdot 10^{-10}$ \cite{WMAP3} 
of the Universe. One of the most popular 
mechanisms to create the baryon asymmetry is leptogenesis \cite{lepto}. 
The heavy neutrinos, whose comparatively huge masses govern the 
smallness of the 
light neutrino masses, decay in the early Universe into Higgs 
bosons and leptons, 
thereby generating a lepton asymmetry, 
which in turn gets recycled into a baryon asymmetry via 
non-perturbative Standard Model processes. 
For recent reviews, see \cite{bumu}. 
In principle, all three heavy neutrinos generate a decay asymmetry, 
which can be written as (summed over all flavors)  
\bea 
\varepsilon_i \D 
= \frac{1}{8 \pi \, v_u^2} \, 
\frac{1}{\left(\tilde{m}_D \, \tilde{m}_D^\dagger\right)_{ii}}  
\, \sum\limits_{j \neq i} 
{\rm Im} \left\{ \left(\tilde{m}_D \, \tilde{m}_D^\dagger 
\right)_{ji}^2 \right\} \, 
\sqrt{x_j} \, \left(
\frac{2}{1 - x_j} - \ln \left( \frac{1+x_j}{x_j} \right) 
 \right) ~,
\\[0.3cm] \D 
\varepsilon_1 \simeq - \frac{3}{8 \pi\, v_u^2} 
\, \frac{1}{\left(\tilde{m}_D \, \tilde{m}_D^\dagger\right)_{11}} 
\, \sum\limits_{j = 2,3} 
\, {\rm Im} \left\{ \left(\tilde{m}_D \, \tilde{m}_D^\dagger 
\right)_{j1}^2 \right\} \, 
\frac{M_1}{M_j}~, 
\label{e1H}
\eea
where  $x_j = M_j^2 /M_i^2$. 
This is the general form of $\varepsilon_i$ and the limit 
for $\varepsilon_1$ in case of $M_3 \gg M_2 \gg M_1$. 
Note that the decay asymmetries depend on 
$\tilde{m}_D \, \tilde{m}_D^\dagger$, which has to be compared to the 
dependence on $\tilde{m}_D^\dagger \, \tilde{m}_D$ governing the  
LFV decays. 
In the case $M_3 \gg M_2 \gg M_1$ only $\varepsilon_1$ 
plays a role, and dedicated numerical studies \cite{bumu,CERN04} have 
shown that in 
case of the MSSM and a hierarchical spectrum of the heavy Majorana 
neutrino masses, successful thermal leptogenesis is only 
possible for
\begin{align}
m_1 \ls 0.1~{\rm eV} \mbox{ and } M_1 \gs 10^{9}~{\rm GeV} ~.
\label{maxtilm1}
\end{align}
However, 
it can occur in certain models that the lightest heavy 
neutrino mass is smaller than the limit of $10^9$ GeV given above. 
We will encounter a scenario like this in the next section. 
There are three possible ways to resolve this problem: 
\begin{itemize}
\item[(i)] the decay of the second heaviest neutrino 
can in certain scenarios generate the baryon asymmetry. 
Flavor effects \cite{flav0,flav1} are important in this respect; 
\item[(ii)] if the heavy Majorana neutrinos are quasi-degenerate in mass, 
the decay asymmetry can be resonantly enhanced, as has been 
analyzed in \cite{YBreso}. This requires some amount of tuning; 
\item[(iii)] non-thermal 
leptogenesis, i.e., the production of heavy neutrinos via inflaton decay
\cite{non_ther}. This possibility is a more model dependent case and 
complicates the situation, as the reheating temperature, the 
mass of the inflaton and the corresponding branching ratios  
for its decay into the Majorana neutrinos need to be known.
\end{itemize}
Let us comment a bit on the first case: 
the expression for the decay asymmetry Eq.~(\ref{e1H}) has been obtained by 
summing over all flavors in which the heavy neutrino decays. 
Recently, however, is has been realized that flavor effects on leptogenesis 
can have significant impact on the scenario \cite{flav0,flav1}. 
The decay asymmetry for the decay of the heavy neutrino in 
a lepton of flavor $\alpha = e, \mu, \tau$ has to be evaluated 
individually and the wash-out or distribution for each flavor 
has to be followed individually by its own Boltzmann-equation. 
However, the bound on the lightest heavy 
neutrino mass $M_1$ is essentially the same as in the ``summed over 
all flavors'' approach. In addition, the decay asymmetry in 
this approach can be enhanced by at most one order of magnitude. 
What will be interesting for our purpose is that if $M_1 \ll 10^9$ GeV 
the second heaviest neutrino with mass $M_2$ can in principle 
generate the baryon asymmetry \cite{flav1}, 
as long as the wash-out by the lightest heavy neutrino 
is low. We will discuss this in more detail in Section \ref{sec:YB}.

\section{\label{sec:first}First Realization of QLC}
The first framework in which our analysis is set is the 
following: 
\begin{itemize}
\item we assume the conventional see-saw mechanism 
to generate the neutrino mass matrix 
$m_\nu = - m_D^T \, M_R^{-1} \, m_D$. Diagonalization 
of $m_\nu$ is achieved via 
$m_\nu = U_\nu^\ast \, m_\nu^{\rm diag} \, U_\nu^\dagger$ and $U_\nu$ 
produces exact bimaximal mixing; 
\item the PMNS matrix is given by $U = U_\ell^\dagger \, U_\nu$, 
where $U_\ell$ corresponds to the CKM matrix $V$. This can be achieved in 
some $SU(5)$ models, in which $m_\ell = m_{\rm down}^T$, where 
$m_{\rm down}$ is the down-quark mass matrix. Hence, 
$V_{\rm down} = V$. 
Consequently, the up-quark mass matrix $m_{\rm up}$ is 
real and diagonal; 
\item in some $SO(10)$ models it holds that $m_{\rm up} = m_D$.  
It follows that the bimaximal structure of $m_\nu$ originates from $M_R$, 
which is diagonalized by $M_R = V_R^\ast \, M_R^{\rm diag} \, V_R^\dagger$.
\end{itemize}
This scenario has been outlined already in \cite{QLC0,QLC1}. 
Note that only $U_\ell = V$ is required for the 
low energy realization of QLC and that the relation 
$m_\ell = m_{\rm down}^T$ will not be required to calculate the 
branching ratios of the LFV decays or the baryon asymmetry. 
It is known that $m_\ell = m_{\rm down}^T$ 
is not realistic for the first and second fermion generation. 
More ``realistic'' 
scenarios have been analyzed in Refs.~\cite{QLC3,LFV_QLC}, in which 
the relation $m_\ell = m_{\rm down}^T$ is modified by the 
Georgi-Jarlskog factor \cite{gj}. However, 
in this case the neutrinos can not be diagonalized by a bimaximal 
mixing matrix, because a too large solar neutrino mixing angle 
would result. Consequently  
the minimality of the scenarios is lost, and the QLC relation 
$\theta_{12} + \theta_C = \pi/4$ turns out to be just 
a numerical coincidence. 
Therefore, following most of the analyzes in 
Refs.~\cite{QLC0,QLC1,QLC2}, we assume that there is a  
particular structure on the mass matrices in which 
mixing depends only weakly on the mass eigenvalues.\\

With the indicated set of properties, we can 
express Eq.~(\ref{eq:pmns_def}) as 
\be
U = V^\dagger \, U_{\rm bimax}~,
\ee
with $U_{\rm bimax}$ corresponding to bimaximal mixing, which will be 
precisely defined in Eq.~(\ref{eq:Ubima}). Moreover, 
Eq.~(\ref{eq:md_bas}) changes to
\be \label{eq:mdt1}
\tilde{m}_D = V_R^T \, m_D \,  V 
\Rightarrow 
\left\{ 
\baz 
\tilde{m}_D^\dagger \, \tilde{m}_D 
=  V^\dagger \, {\rm diag}(m_u^2, m_c^2, m_t^2) \, V
& \mbox{ for LFV}~,\\[0.3cm]
\tilde{m}_D \, \tilde{m}_D^\dagger 
= V_R^T \, {\rm diag}(m_u^2, m_c^2, m_t^2) \,  V_R^\ast 
&  \mbox{ for } \eta_B
~.
\ea 
\right. 
\ee
In the above equation we have given the two important matrices  
$\tilde{m}_D \, \tilde{m}_D^\dagger$ and $\tilde{m}_D^\dagger \, \tilde{m}_D$  
describing leptogenesis and the branching ratios of the 
lepton flavor violating processes. Note however, that 
for the latter we have for now neglected the 
logarithmic dependence on the heavy 
neutrino masses, cf.~Eq.~(\ref{eq_ijg}).

\subsection{\label{sec:low}Low Energy Neutrino Phenomenology}

The matrix diagonalizing $m_\nu$ is called $U_\nu$ and 
corresponds to a bimaximal mixing matrix: 
\be
\label{eq:Ubima}
U_\nu = U_{\rm bimax} = 
P_\nu \, \tilde{U}_{\rm bimax} \, Q_\nu = 
{\rm diag}(1,e^{i \phi}, e^{i \omega}) \, 
\left(
\bad  
\frac{1}{\sqrt{2}} &  \frac{1}{\sqrt{2}} &  0 \\[0.3cm]
-\frac{1}{2} &  \frac{1}{2} &  \frac{1}{\sqrt{2}} \\[0.3cm]
\frac{1}{2} &  -\frac{1}{2} &  \frac{1}{\sqrt{2}} \\[0.3cm]
\ea 
\right){\rm diag}(1,e^{i \sigma}, e^{i \tau}) ~.
\ee
We have included two diagonal phase matrices $P_\nu$ and $Q_\nu$. 
It has been shown in Ref.~\cite{FPR} that this is the 
most general form if all "unphysical" phases are rotated away. 
We have in total five phases, one phase in $U_\ell = V$ 
and four phases in $U_\nu$. 
Note that $Q_\nu$ is ``Majorana-like'' \cite{FPR}, i.e., the phases 
$\sigma$ and $\tau$ do not appear in neutrino oscillations, 
but contribute to the low energy Majorana phases. 
Multiplying the matrices of Eq.~(\ref{eq:CKM}) and Eq.~(\ref{eq:Ubima}) yields 
for the oscillation parameters: 
\bea \label{eq:QLCobs}
U = U_\ell^\dagger \, U_\nu = V^\dagger \, U_{\rm bimax} 
\\[0.3cm] \hspace{-1.3cm}
\Rightarrow 
\left\{ 
\ba
\sin^2 \theta_{12} = \frac{\D 1}{\D 2} 
- \frac{\D \lambda }{\D \sqrt{2}} \, \cos \phi 
+ {\cal O}(\lambda^3) ~,\\[0.2cm]
|U_{e3}| = \frac{\D \lambda}{\D \sqrt{2}} + {\cal O}(\lambda^3) ~,\\[0.2cm]
\sin^2 \theta_{23} = \frac{\D 1}{\D 2} - 
\left(
A \, \cos (\omega - \phi) + \frac 14
\right) \, \lambda^2 + {\cal O}(\lambda^4)~,\\[0.2cm]
J_{CP}^{\rm lep} = \frac{\D \lambda}{\D 4\sqrt{2}} \, \sin \phi 
+ {\cal O}(\lambda^3)~.
\ea 
\right. 
\eea
Apparently, Eq.~(\ref{eq:QLCobs}) generates correlations between 
the observables.   
The solar neutrino mixing parameter depends on the $CP$ phase $\phi$, 
which originates from the neutrino sector 
and is to a very good approximation the phase governing leptonic 
$CP$ violation in oscillation experiments. 
Note that in order to have solar neutrino mixing of the observed 
magnitude, the phase has to be close to zero or 
$2\pi$. Approximately, at $3\sigma$ it should be 
below $\pi/4$ or above $7\pi/4$. 
The smallest solar neutrino mixing angle is obtained for $\phi = 0$ and 
the prediction for $\sin^2 \theta_{12}$ is 
\be
\sin^2 \theta_{12} \gs 0.334~(0.333, \, 0.332, \, 0.331)~. 
\ee
This value of $\sin^2 \theta_{12} \gs 0.33$ has to be compared with 
the experimental $1\sigma$ ($2\sigma$) limit 
of $\sin^2 \theta_{12} \le 0.33~(0.37)$, 
showing a small conflict.
Note that for the numerical values, as well as for the generation 
of the plots, which will be presented and discussed in the following, we 
did not use the approximate expressions in 
Eq.~(\ref{eq:QLCobs}), but the exact formulae\footnote{ 
Note for instance that the next term in the expansion of 
$|U_{e3}|$ is of order 
$\lambda^3 \simeq 0.01$ and can contribute sizably.}. 
Besides the phases, 
we also vary the parameters of the CKM matrix in their 1, 2 and 3$\sigma$ 
ranges (though in particular the error in $\lambda$ is negligible), and also 
fix these parameters to their best-fit values. 
Even for the best-fit values of the CKM parameters, 
there results a range of values, which is caused by the 
presence of the unknown phases $\phi$ and $\omega$. 
To a good approximation, $|U_{e3}|$ is the sine of the 
Cabibbo angle divided by $\sqrt{2}$, leading to a sharp prediction of 
$|U_{e3}|^2 = 0.0258$. 
Varying the phases and the CKM parameters, we find a range of 
\be
|U_{e3}| = 0.1607^{+0.0058, \, 0.0069, \, 0.0083, \, 0.0096}
_{-0.0059, \, 0.0068, \, 0.0080, \, 0.0091}
~,
\ee
where we took the central value $\lambda/\sqrt{2} = 0.1607$. 
Recall that the $1\sigma$ ($2\sigma$) bound on $|U_{e3}|$ is 0.11 (0.17). 
Therefore, the prediction for $|U_{e3}|$ is incompatible 
with the current 1$\sigma$ bound of $|U_{e3}|$ and even quite 
close to the $2\sigma$ limit. 
The experiments taking data in the next 5 to 10 years \cite{exp_new} 
will have to 
find a signal corresponding to non-vanishing  $|U_{e3}|$ in order for this 
particular framework to survive.
Leptonic $CP$ violation is in leading order proportional to 
$\lambda \, \sin \phi$, which is five orders in units of $\lambda$ larger 
than the $J_{CP}$ of the quark sector. 
If the neutrino sector conserved $CP$, one would obtain 
$J_{CP}^{\rm lep} = \frac 18 \, A \, \eta \, \lambda^4$, which is still 
two orders of $\lambda$ larger than the $J_{CP}$ of the quark sector. 
If $V$ was equal to the unit matrix, which corresponds to 
bimaximal mixing in the PMNS matrix, $J_{CP}^{\rm lep}$ would be zero. 
There is an interesting ``sum-rule'' between leptonic $CP$ violation, 
solar neutrino mixing and $|U_{e3}|$: 
\bea \label{eq:sumrule}
\sin^2 \theta_{12} \simeq \frac 12 - |U_{e3}| \, \cos \phi \simeq 
\frac 12 \pm \sqrt{|U_{e3}|^2 - 16 \, (J_{CP}^{\rm lep})^2} 
~.
\eea
Overall, the experimental result of $\sin^2 \theta_{12} \simeq 0.31$ implies 
large $\cos \phi$, and therefore small $\sin \phi$, leading to 
small $CP$ violating effects even though $|U_{e3}|$ is sizable. 
Atmospheric neutrino mixing stays very close to maximal
and due to cancellations $\sin^2 \theta_{23} = \frac 12$ can always 
occur. 
If $\cos(\omega - \phi) = 1$, then $\sin^2 \theta_{23}$ takes its 
minimal value. We have seen above that the observed low value of the 
solar neutrino mixing angle requires $\phi \simeq 0$, so that 
$\omega \simeq 0$ is implied when $\theta_{23}$ is very close to maximal. 
The minimal and maximal values of $\sin^2 \theta_{23}$ are 
given by 
\be
\sin^2 \theta_{23} \ge 0.445~(0.444, \, 0.443, \, 0.442)~\mbox{ and } 
\sin^2 \theta_{23} \le 0.531~(0.532, \, 0.533, \, 0.534)~.
\ee
Probing deviations from maximal mixing of order 10\% could be 
possible in future experiments \cite{exp_new}. 
In Fig.~\ref{fig:the_one_ring} we show the correlations between the 
oscillation parameters which result from the relation 
$U = V^\dagger \, U_{\rm bimax}$ 
in Eq.~(\ref{eq:QLCobs}). 
We plot $J_{CP}^{\rm lep}$, 
$\phi$ and $\sin^2 \theta_{23}$ against $\sin^2 \theta_{12}$, 
as well as $\sin^2 \theta_{23}$ against $|U_{e3}|$. 
We also indicate the current 
1, 2 and $3\sigma$ ranges of the oscillation parameters. 
This shows again that solar neutrino mixing is predicted to be 
close to its 1$\sigma$ bound 
and $|U_{e3}|$ even close to its 2$\sigma$ bound.\\

Now we turn to the neutrino observables outside the oscillation 
framework and comment on the consequences for neutrinoless double beta decay. 
The two invariants related to the Majorana phases are 
\bea \label{eq:S121}
S_1 = \frac{\D \lambda}{\D 2} \, \sin (\phi + \tau) 
+ \frac{\D \lambda^2}{\D 2\sqrt{2}} \, \sin \tau + {\cal O}(\lambda^3)
~,\\[0.3cm]
S_2 = \frac{\D \lambda}{\D 2} \, \sin (\phi - \sigma + \tau)  
+ \frac{\D \lambda^2}{\D 2\sqrt{2}} \, 
\sin (\sigma - \tau) + {\cal O}(\lambda^3)
~.
\eea
As expected, the two phases $\sigma$ and $\tau$ 
in $Q_\nu$ only appear in these quantities. According to the 
parameterization of Eq.~(\ref{eq:Upara}), we have
$S_1 = -c_{12} \, c_{13} \, s_{13} \, s_\beta$ and 
$S_2 = s_{12} \, c_{13} \, s_{13} \, s_{\alpha - \beta}$. We can insert 
in Eq.~(\ref{eq:S121}) the expressions 
for the mixing angles from Eq.~(\ref{eq:QLCobs}) to obtain 
in leading order $\sin \beta \simeq - \sin (\phi + \tau)$ and 
$\sin (\alpha - \beta) \simeq \sin (\phi - \sigma + \tau)$. Hence, 
the Majorana phase $\sigma$ is related to the phase $\alpha$ in the 
parameterization of Eq.~(\ref{eq:Upara}). 
It is interesting to study the form of the neutrino mass matrix, which 
is responsible for bimaximal mixing. 
It reads 
\bea \label{eq:mnubimax}
m_\nu^{\rm bimax} = 
\left( 
\bad 
\D A & B \, e^{-i \phi}  & - B \, e^{-i \omega}  \\[0.2cm] 
\D \cdot &  (D + \frac{A}{2}) \, e^{-2i \phi} 
&  (D - \frac{A}{2}) \, e^{-i (\phi + \omega)} \\[0.3cm]
\D \cdot & \cdot & (D + \frac{A}{2} ) \, e^{-2i \omega}
\ea   
\right) \\[0.3cm]
= 
\left( 
\bad 
1 & 0 & 0 \\[0.2cm]
0 & e^{-i \phi} & 0 \\[0.2cm]
0 & 0 & e^{-i \omega} 
\ea 
\right) 
\left( 
\bad 
\D A & B   & - B   \\[0.2cm] 
\D \cdot &  D + \frac{A}{2}
&  D - \frac{A}{2}  \\[0.3cm]
\D \cdot & \cdot & D + \frac{A}{2} 
\ea   
\right) 
\left( 
\bad 
1 & 0 & 0 \\[0.2cm]
0 & e^{-i \phi} & 0 \\[0.2cm]
0 & 0 & e^{-i \omega} 
\ea 
\right) ~, 
\eea
where 
\be \label{eq:ABD}
A = \frac 12 \left( m_1 + m_2 \, e^{-2 i \sigma}\right) ~,~
B = \frac{1}{2 \sqrt{2}} \left( m_2 \, e^{-2 i \sigma} - m_1 \right)~
,~D = \frac{m_3 \, e^{-2 i \tau}}{2}~. 
\ee
The inner matrix in Eq.~(\ref{eq:mnubimax}) is diagonalized by 
a real and bimaximal rotation and the masses are obtained as 
\be \label{eq:massABD}
m_1 = A - \sqrt{2} \, B~~,~ e^{-2 i \sigma} \, m_2 = A + \sqrt{2} \, B~~,~ 
e^{-2 i \tau} \, m_3 = 2 \, D~.
\ee

Up to now there has been no need to specify the neutrino mass ordering. 
This is however necessary in order to 
discuss \onbb (\obb) \cite{0vbbrev}. 
There are three extreme hierarchies often discussed; the normal 
hierarchy ($m_3 \simeq \sqrt{\dma} \gg m_{2} \simeq \sqrt{\dms} \gg m_1$),  
the inverted hierarchy 
($m_2 \simeq m_1 \simeq \sqrt{\dma} \gg m_{3}$) and the 
quasi-degenerate case ($m_1 \simeq m_2 \simeq m_3  \gg \sqrt{\dma}$). 
The effective mass which can be measured in \obb~experiments 
is the $ee$ element of $m_\nu$ 
in the charged lepton basis. To first order in $\lambda$ one gets 
for a normal hierarchy that 
$\meff \simeq \frac 12 \, \sqrt{\dms} \, \lambda$. 
In case of an inverted hierarchy we have  
\be \label{eq:meff1} 
\meff \simeq 
\sqrt{\dma} \, \left| 
c_\sigma + \sqrt{2} \, s_\sigma \, s_{\phi} \, \lambda 
\right| 
~.
\ee
The maximal (minimal) effective mass is obtained for 
$\sigma = 0$ ($\sigma = \pi/2$). On the other hand, 
 we have 
$\meff \simeq \sqrt{\dma} \, 
\sqrt{1 - \sin^2 2 \theta_{12} \, \sin^2 \alpha}$ in 
terms of the usual parameterization \cite{0vbbrev}. Therefore, 
as is also obvious from the discussion following Eq.~(\ref{eq:S121}), 
$\sigma$ will be closely related to the Majorana phase $\alpha$. 
Similar considerations apply to the quasi-degenerate case.

\subsection{\label{sec:LFV}Lepton Flavor Violation}
Now we study the branching ratios of the LFV decays like $\mu \ra e \gamma$ 
for this scenario. 
With our present assumptions we have that $m_D = m_{\rm up} = 
{\rm diag}(m_u, m_c, m_t)$.  
With this input and with Eq.~(\ref{eq:mdt1}) one easily obtains 
\be \label{eq:mueg}
\left| (\tilde{m}_D^\dagger  \, \tilde{m}_D )_{21}
\right|^2 \simeq A^4 \, m_t^4 
\left(\eta^2 - (1 - \rho)^2 \right) \, \lambda^{10} + 
{\cal O}(\lambda^{14})~.
\ee
Note that we have neglected the logarithmic dependence on $M_i$. 
The double ratios are\footnote{The relative magnitude 
of the branching ratios has 
in this scenario been estimated in Ref.~\cite{LFV_QLC}. Here we 
take the dependence on $M_i$ and $m_i$ carefully into account and 
study in addition their absolute magnitude.}
\be \label{eq:double}
R (21/31) \simeq A^2 \, \lambda^4 + 
{\cal O}(\lambda^{8})~~,~~
R (21/32) \simeq A^2 \, \left(\eta^2 - (1 - \rho)^2 \right) \, \lambda^6
+ {\cal O}(\lambda^{10})~.
\ee
The branching ratios behave according to 
\be \label{eq:behave}
\text{BR}(\mu \ra e \gamma):\text{BR}(\tau \ra e \gamma):
\text{BR}(\tau \ra \mu \gamma) \simeq \lambda^6 : \lambda^2 : 1~,
\ee
which is in agreement with Ref.\ \cite{LFV_QLC}.\\

In order to conduct a more precise study of the rates of the LFV processes, 
we recall that there is some dependence on the heavy neutrino masses, as 
encoded in the matrix 
$L = \delta_{ij} \, \ln M_X/M_i$ in Eq.~(\ref{eq_ijg}). 
Hence, we need to evaluate the values of the heavy Majorana neutrino 
masses, i.e., we need to invert the see-saw formula 
$m_\nu = -m_D^T \, M_R^{-1} \, m_D$ 
and diagonalize $M_R$ \cite{others,AFS0,AFS}.  
The light neutrino mass matrix is displayed in 
Eq.~(\ref{eq:mnubimax}). 
With $m_D = m_{\rm up} = {\rm diag}(m_u, m_c, m_t)$ the heavy 
neutrino mass matrix reads: 
\bea \hspace{-1cm} \label{eq:MRbimax}
-M_R = m_{\rm up} \, m_\nu^{-1} \, m_{\rm up} = 
P_\nu \, 
\left( 
\bad 
\tilde{A} \, m_u^2 & \tilde{B} \, m_u \, m_c 
& -\tilde{B} \, m_u \, m_t \\[0.2cm]
\cdot & 
\left(\tilde{D} + \frac{\tilde{A}}{2}\right) \, m_c^2 &  
\left(\tilde{D} - \frac{\tilde{A}}{2}\right) \, m_c \, m_t   \\[0.2cm]
\cdot & \cdot & 
\left(\tilde{D} + \frac{\tilde{A}}{2}\right) \, m_t^2 
\ea 
\right)
\, P_\nu  
~,
\eea
where 
\[ 
\tilde{A} = \frac{1}{2 \, m_1} + \frac{e^{2i \sigma}}{2 \, m_2} = 
\frac{A}{A^2 - 2 B^2}~~,~
\tilde{B} = 
\frac{e^{2i \sigma}}{2 \sqrt{2} \, m_2} - \frac{1}{2 \sqrt{2} \, m_1} 
= \frac{-B}{A^2 - 2 B^2} ~~,~
\tilde{D} = \frac{e^{2 i \tau}}{2 \, m_3} = 
\frac{1}{4 D}~.
\]
$A$, $B$ and $D$ are given in Eq.~(\ref{eq:ABD}). 
The heavy Majorana mass matrix is related to the inverse of the 
light neutrino mass matrix and has for bimaximal mixing a very similar form. 
Due to the very hierarchical structure of $M_R$, 
and if none of the elements vanish, 
the eigenvalues are quite easy to obtain (see also \cite{AFS}): 
\bea \label{eq:N1}
\D M_1 \, e^{i \phi_1} \simeq 
\frac{m_u^2 \, (\tilde A^2 - 2 \, \tilde B^2)}{\tilde A} 
= \frac{2 \, m_u^2}{m_1 + m_2 \, e^{-2 i \sigma}}
~~,~~\\[0.3cm]\D 
M_2 \, e^{i \phi_2} \simeq 2 \, m_c^2 \frac{\tilde A \, \tilde D}{\tilde D + 
\tilde A /2}
= 2 \, e^{2i(\sigma + \tau)} \, m_c^2 \, 
\frac{m_1 + m_2 \,e^{-2 i \sigma} }{m_2 \, m_3 + m_1 \, m_3 \,e^{2 i \sigma} 
+ 2 \, m_1 \, m_2 \, e^{2 i \tau}}
~~,~~\\[0.3cm]\D 
M_3\, e^{i \phi_3} \simeq  m_t^2 \, (\tilde D + \tilde A /2 )  = 
\frac{m_t^2}{4 \, m_1 \, m_2 \, m_3} \, 
\left( 
2 e^{2 i \tau} m_1 \, m_2 + e^{2 i \sigma} \, m_1 \, m_3 + m_2 \, m_3
\right)~.
\eea
Here $M_{1,2,3}$ are real and positive,  
and $\phi_{1,2,3}$ denote the phases of the 
complex eigenvalues of the inner matrix in Eq.~(\ref{eq:MRbimax}). 
We see that the values of the heavy Majorana masses depend on the 
phases $\sigma$ and $\tau$, which in turn are related to the low 
energy Majorana phases. 
Note that the requirement of $M_3$ from Eq.~(\ref{eq:N1}) 
being smaller than the Planck mass 
gives a lower bound on the smallest neutrino mass of 
\be \label{eq:m11} 
m_1 \ge \frac{m_t^2}{4 \, M_{\rm Pl}} \simeq 2 \cdot 10^{-7}~{\rm eV }
\mbox{ and } 
m_3 \ge \frac{m_t^2}{2 \, M_{\rm Pl}} \simeq 4 \cdot 
10^{-7}~{\rm eV}~, 
\ee
for the normal and inverted hierarchy, respectively.\\

\begin{figure}[htb]\hspace{-1cm}
\epsfig{file=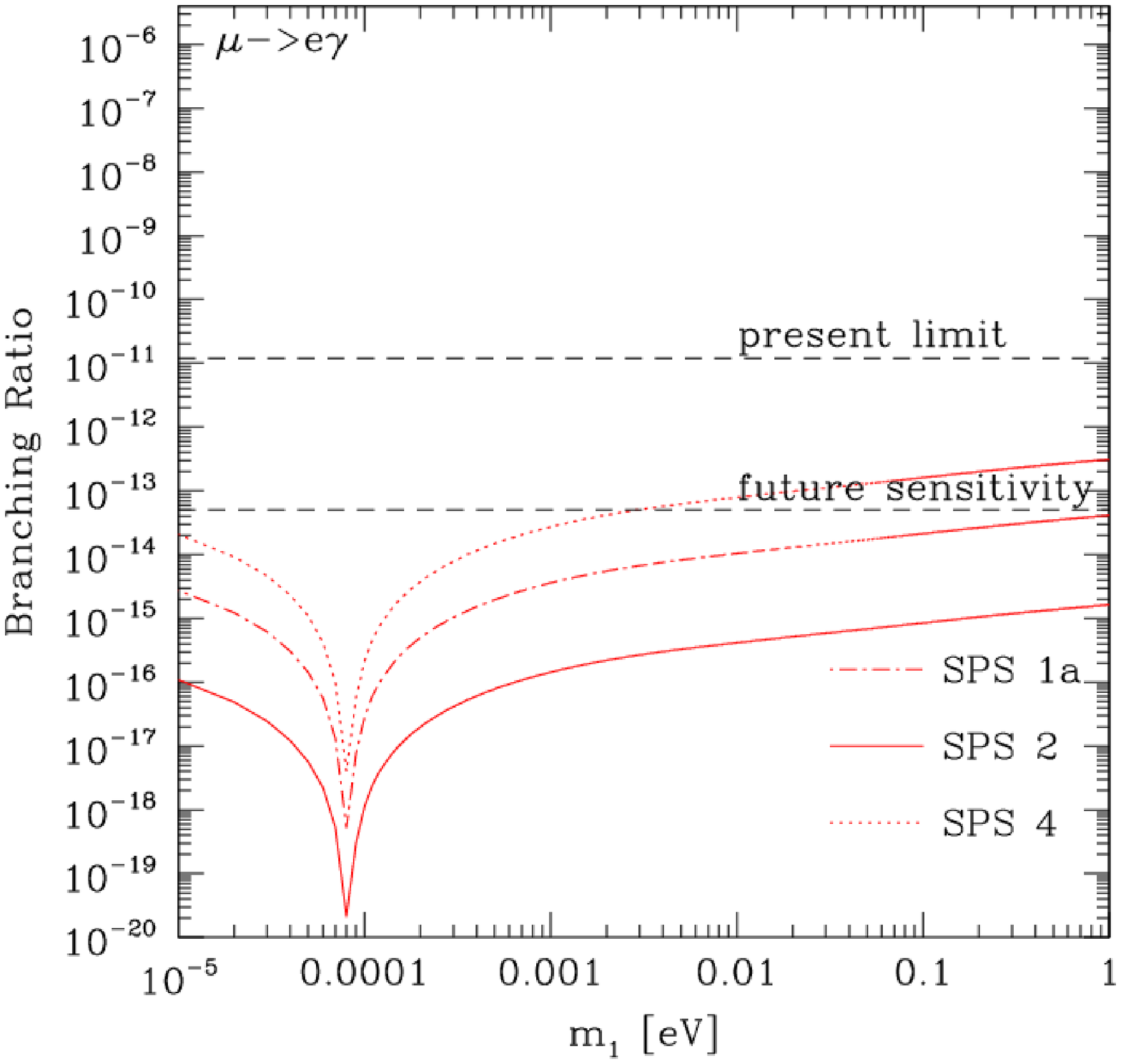,width=5.6cm,height=8cm}
\epsfig{file=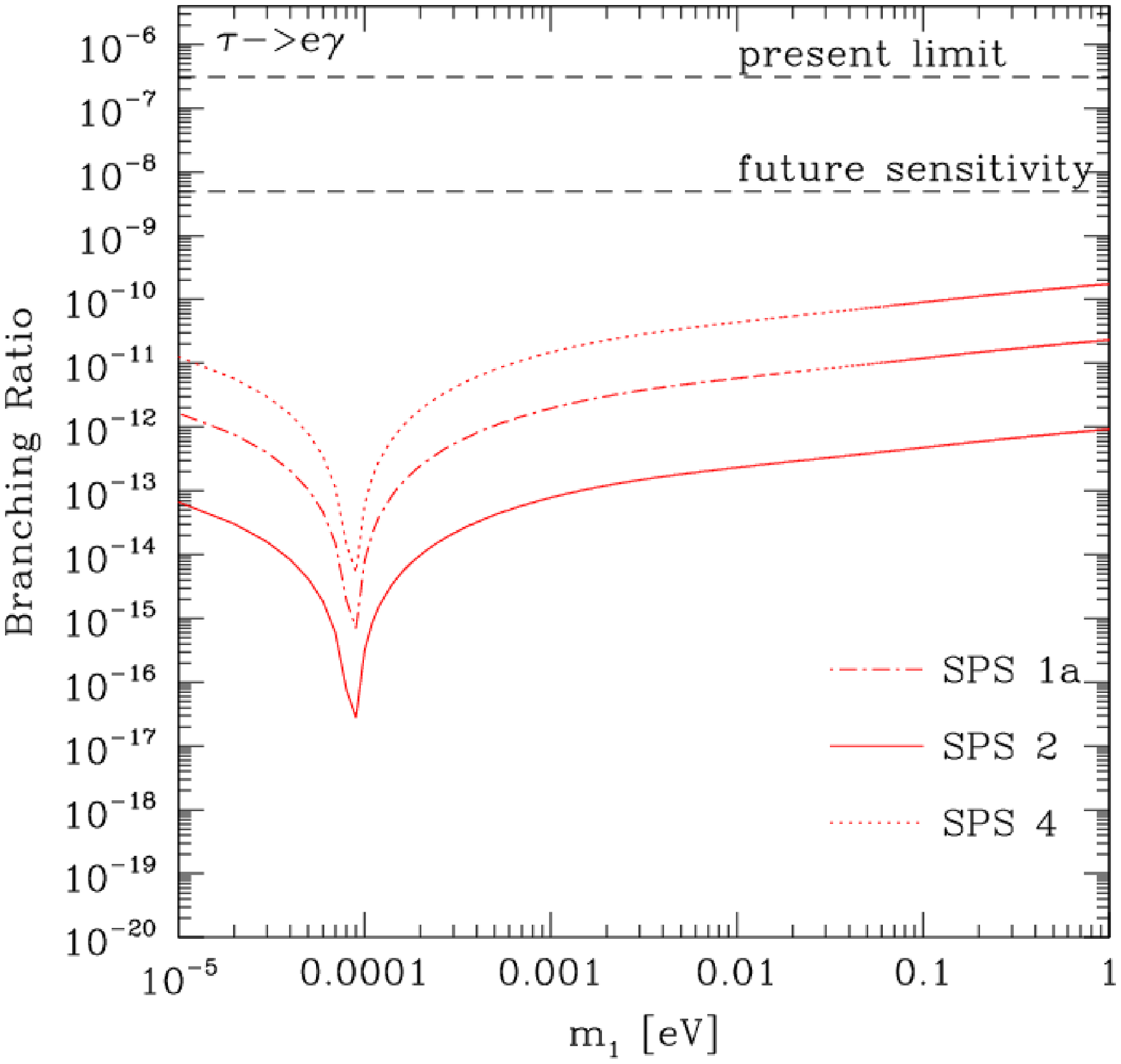,width=5.6cm,height=8cm}
\epsfig{file=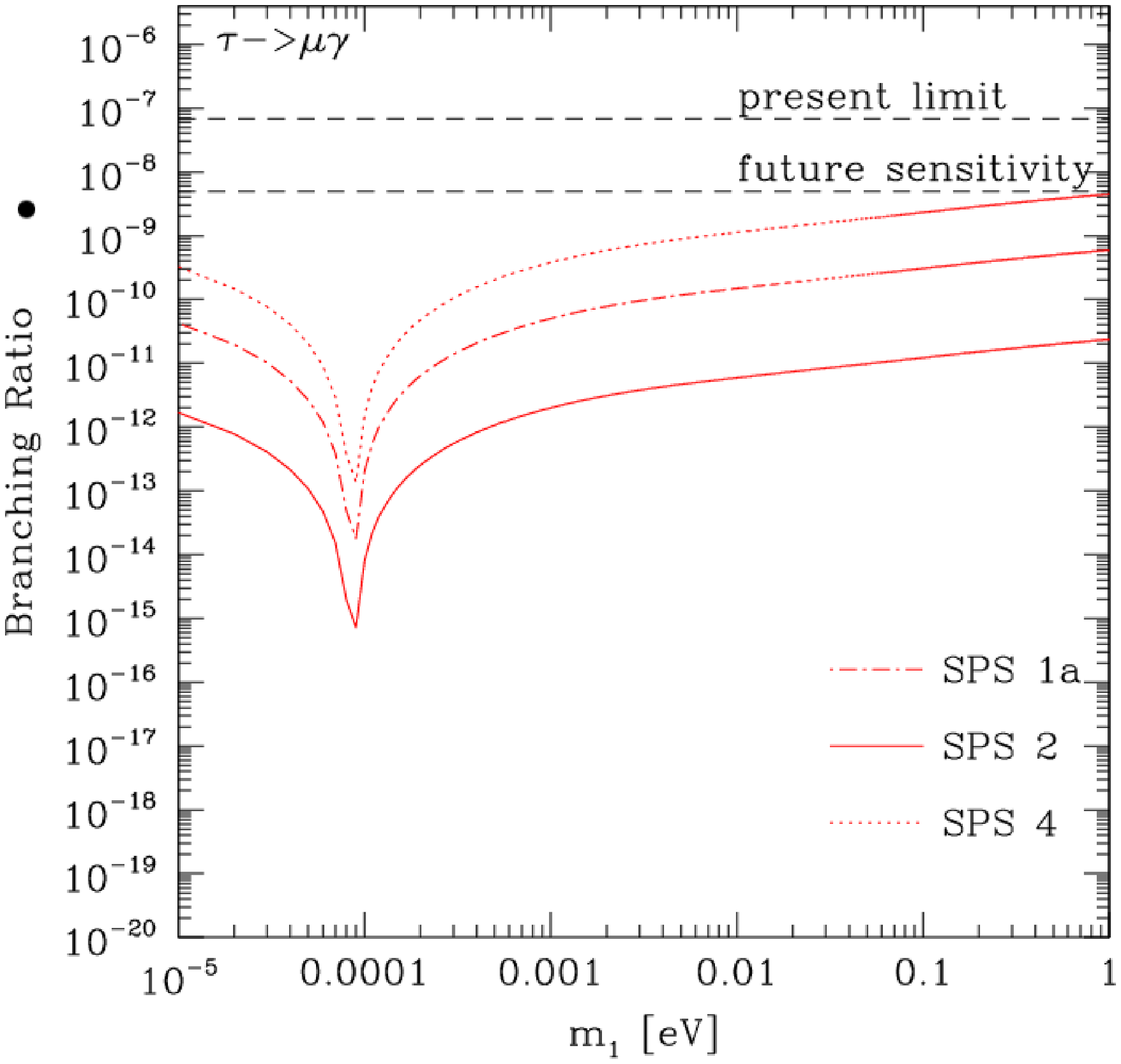,width=5.6cm,height=8cm}
\caption{\label{fig:BR1NH}First realization of QLC: 
the branching ratios for $\mu \ra e \gamma$, 
$\tau \ra e \gamma$ and $\tau \ra \mu \gamma$ against the smallest 
neutrino mass (normal ordering) 
for the SPS points 1a, 2 and 4, see Table \ref{tab:SPS}. 
Indicated are also the present and future experimental sensitivities.}
\end{figure}

\begin{figure}[htb]\hspace{-1cm}
\epsfig{file=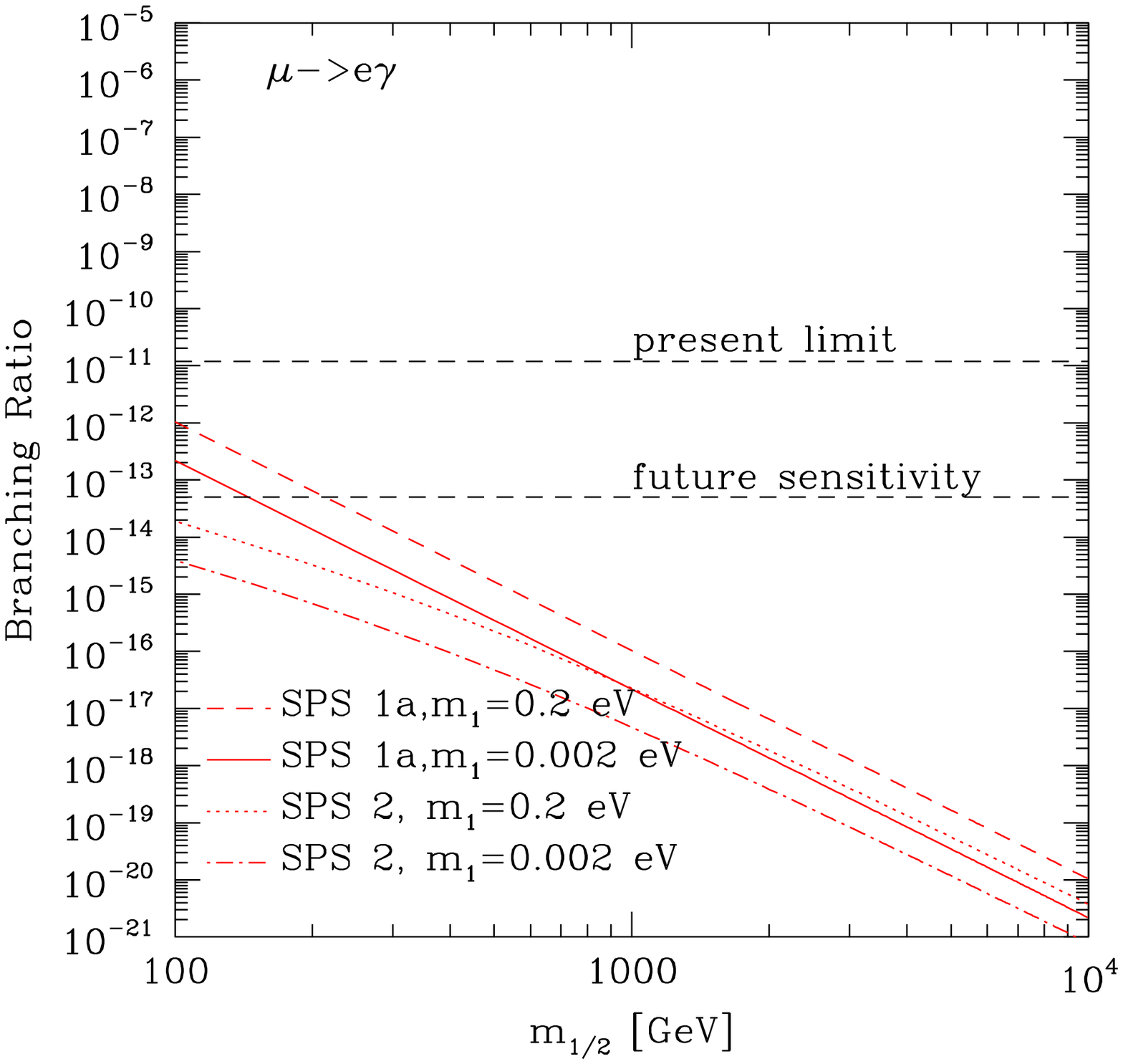,width=5.6cm,height=8cm}
\epsfig{file=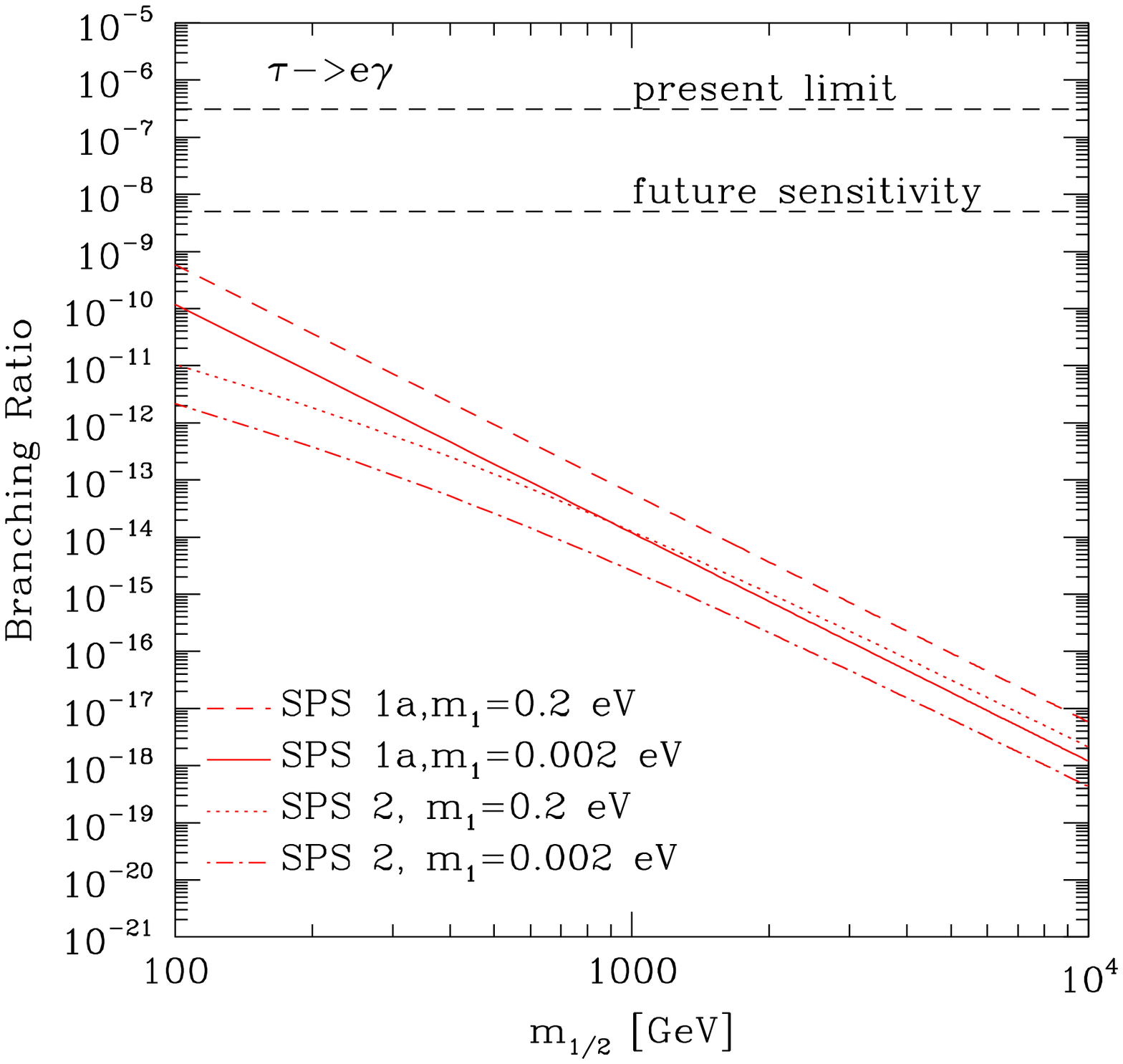,width=5.6cm,height=8cm}
\epsfig{file=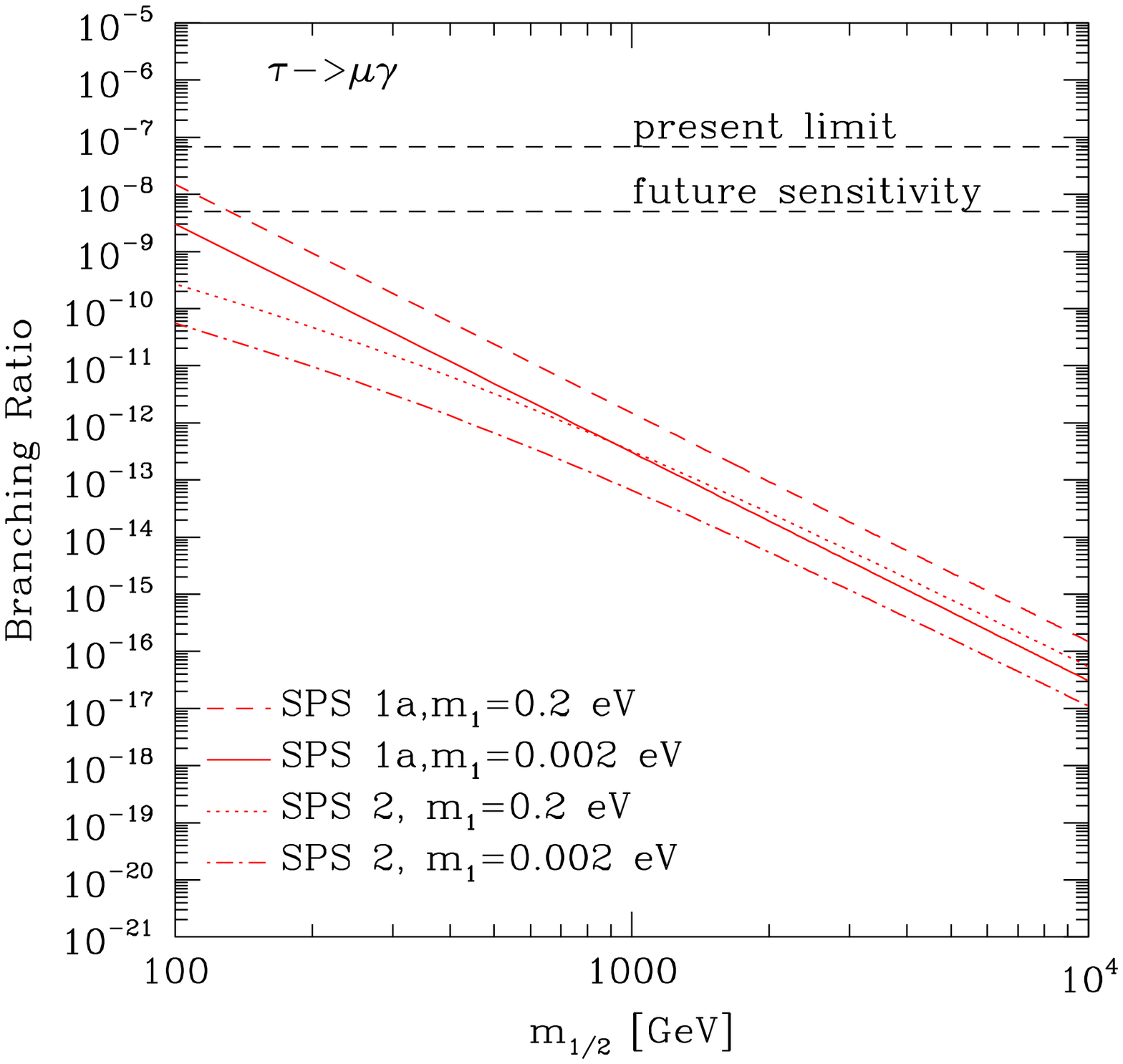,width=5.6cm,height=8cm}
\caption{\label{fig:BRs}First realization of QLC: 
the branching ratios for $\mu \ra e \gamma$, 
$\tau \ra e \gamma$ and $\tau \ra \mu \gamma$ against 
the SUSY parameter $m_{1/2}$ for the SPS slopes 1a and 2 
see Table \ref{tab:SPS}. 
We have chosen two values for the neutrino masses (normal ordering), 
namely 0.002 eV and 0.2 eV. 
Indicated are also the present and future experimental sensitivities.}
\end{figure}

The matrix $V_R$ is defined via 
$M_R = V_R^\ast \, M_R^{\rm diag} \, V_R^\dagger$, where 
$ M_R^{\rm diag} = {\rm diag}(M_1, M_2, M_3)$ contains real and 
positive entries. We find 
\bea \label{eq:VR}
V_R = i \, P_\nu^\ast \, \tilde{V}_R \, 
P_\nu \, R_\nu~,\mbox{ where } \\[0.3cm]
\tilde{V}_R \simeq 
{\scriptsize
\left( 
\bad \D 
1 & \D \frac{m_u}{m_c} \, \frac{\tilde B}{\tilde A} & \D 
-\frac{m_u}{m_t} \, \frac{2\tilde B (\tilde A^2 - 2 \tilde B^2)}
{\tilde A (\tilde A^2 - 2 \tilde B^2 + 2 \tilde A \, \tilde D) 
- 4 \tilde B^2 \, \tilde D} \\[0.2cm]
\D - \frac{m_u}{m_c} \, \frac{\tilde B}{\tilde A} & \D 1 
& \D - \frac{m_c}{m_t} \, 
\frac{\tilde A (\tilde A^2 - 2 \tilde B^2 - 2 \tilde A \, \tilde D) + 
4 \tilde B^2 \, \tilde D}
{\tilde A (\tilde A^2 - 2 \tilde B^2 + 2 \tilde A \, \tilde D) 
- 4 \tilde B^2 \, \tilde D}
\\[0.2cm]
\D \frac{m_u}{m_t} \, \frac{\tilde B}{\tilde A} 
& \D \frac{m_c}{m_t} \, 
\frac{\tilde A (\tilde A^2 - 2 \tilde B^2 - 2 \tilde A \, \tilde D) + 
4 \tilde B^2 \, \tilde D}
{\tilde A (\tilde A^2 - 2 \tilde B^2 + 2 \tilde A \, \tilde D) 
- 4 \tilde B^2 \, \tilde D}
& 1
\ea
\right)~,}
\eea
where $R_\nu = {\rm diag}(e^{-i \phi_1/2}, e^{-i \phi_2/2},e^{-i \phi_3/2})$ 
contains the phases of the eigenvalues in Eq.~(\ref{eq:N1}). 
The above matrix is unitary to order $m_u/m_c$ or $m_c/m_t$, 
which phenomenologically corresponds to an order of $\lambda^4$. 
The heavy neutrino masses are plotted in Fig.\ \ref{fig:N1} as 
a function of the lightest neutrino mass in case of normal ordering. 
Figure \ref{fig:N3} shows the same for inversely ordered light neutrinos. 
We have chosen four different pairs of values for $\sigma$ and $\tau$.
For the plots we have fixed \dms~and \dma~to their best-fit values 
and have taken  
the quark masses as\footnote{The values 
for the heavy neutrino masses are not much different when 
we take the quark masses \cite{high} at a higher energy scale.} 
$m_u = 0.45$ MeV, $m_c = 1.2$ GeV and $m_t = 175 $ GeV. 
The matrix $M_R$ was diagonalized numerically.  
Eq.~(\ref{eq:N1}) is nevertheless an excellent approximation if 
$\sigma$ and $\tau$ are far away from $\pi/2$. 
Moreover, it holds that $V_R \simeq \mathbbm{1}$ in this case.  
On the other 
hand, if $\sigma \simeq \pi/2$ it can occur that $M_1$ and $M_2$ 
are almost degenerate if $m_1$ takes a value around 0.5 eV. 
This happens if $\tilde A=0$ or, 
strictly speaking, $\tilde A \, m_u^2 \ll \tilde{B} \, m_u \, m_c$
in which case Eqs.~(\ref{eq:N1}, \ref{eq:VR}) are no 
longer valid \cite{AFS0,AFS}, but $M_1$ and $M_2$ build 
a pseudo-Dirac pair with mass 
\be \label{eq:PD}
M_1 \simeq M_2 \simeq \tilde B \, m_u \, m_c \simeq 
\frac{m_u \, m_c}{2\sqrt{2} \, m_1} \sim  10^6~\rm GeV~.
\ee
Note that the indicated value of $m_1$ is in conflict with tight 
cosmological constraints \cite{steen}.
There are similar situations for $M_2$ and $M_3$, which occur 
when $\tau \simeq \pi/2$.    
Neglecting these tuned cases, we plot the branching 
ratios in case of $\tau = \sigma = 0$ for the normal ordering in Fig.\ 
\ref{fig:BR1NH}
as a function of the smallest neutrino mass\footnote{Note that 
for inverse mass ordering the masses $m_1$ and $m_2$ are always rather close. 
As obvious from Eq.~(\ref{eq:N1}), this leads to
slightly larger masses for the heavy neutrinos. 
This translates into branching ratios which for small  
$m_3$ are larger by a factor of roughly 3.}, 
choosing the SPS points 1a, 2 and 4. 
We do not use points 1b and 3, because the corresponding 
plots will be indistinguishable from the plots for points 1a and 2, 
respectively.
The results are typical if both 
$\tau$ and $\sigma$ are not close to $\pi/2$. 
In order to take 
renormalization aspects into account, we evaluated the branching ratios 
for quark masses at high scale \cite{high}. For instance, 
we took 
$m_u = 0.7$ MeV, $m_c = 210$ MeV and $m_t = 82.4$ GeV, which corresponds 
for $\tan \beta = 10$ to 
$m_u = 2.3$ MeV, $m_c = 677$ MeV and $m_t = 181$ GeV at $M_Z$ \cite{high}. 
Due to the presence of the diagonal matrix 
$L = \delta_{ij} \, \ln M_X/M_i$ in the equation for the branching ratios 
the possibility of cancellations arises, leading to a very small 
branching ratio. From Eq.~(\ref{eq:mueg}) alone, such a cancellation 
is impossible. We have also indicated current and future 
sensitivities on the decays in Fig.~\ref{fig:BR1NH}. 
Typically, $\mu \ra e \gamma$ can be observable for not too 
small neutrino masses, unless the SUSY masses approach the TeV scale. 
BR($\tau \ra e \gamma$) is predicted to be very small, and 
observation of $\tau \ra \mu \gamma$ requires rather 
large neutrino masses, small SUSY masses, or large $\tan \beta$. 
This is illustrated in Fig.~\ref{fig:BRs}, where we have plotted 
the branching ratios as a function of  
the SUSY parameter $m_{1/2}$ for the SPS slopes 1a and 2 from 
Table \ref{tab:SPS}. 
We have chosen two values for the neutrino masses (normal ordering), 
namely 0.002 eV and 0.2 eV. 
The relative magnitude of the branching ratios, 
as estimated in Eq.~(\ref{eq:behave}), holds true for most of the parameter 
space.

\subsection{\label{sec:YB}Comments on Leptogenesis}
\begin{figure}[htb]\hspace{-1cm}
\begin{tabular}{cc}
\epsfig{file=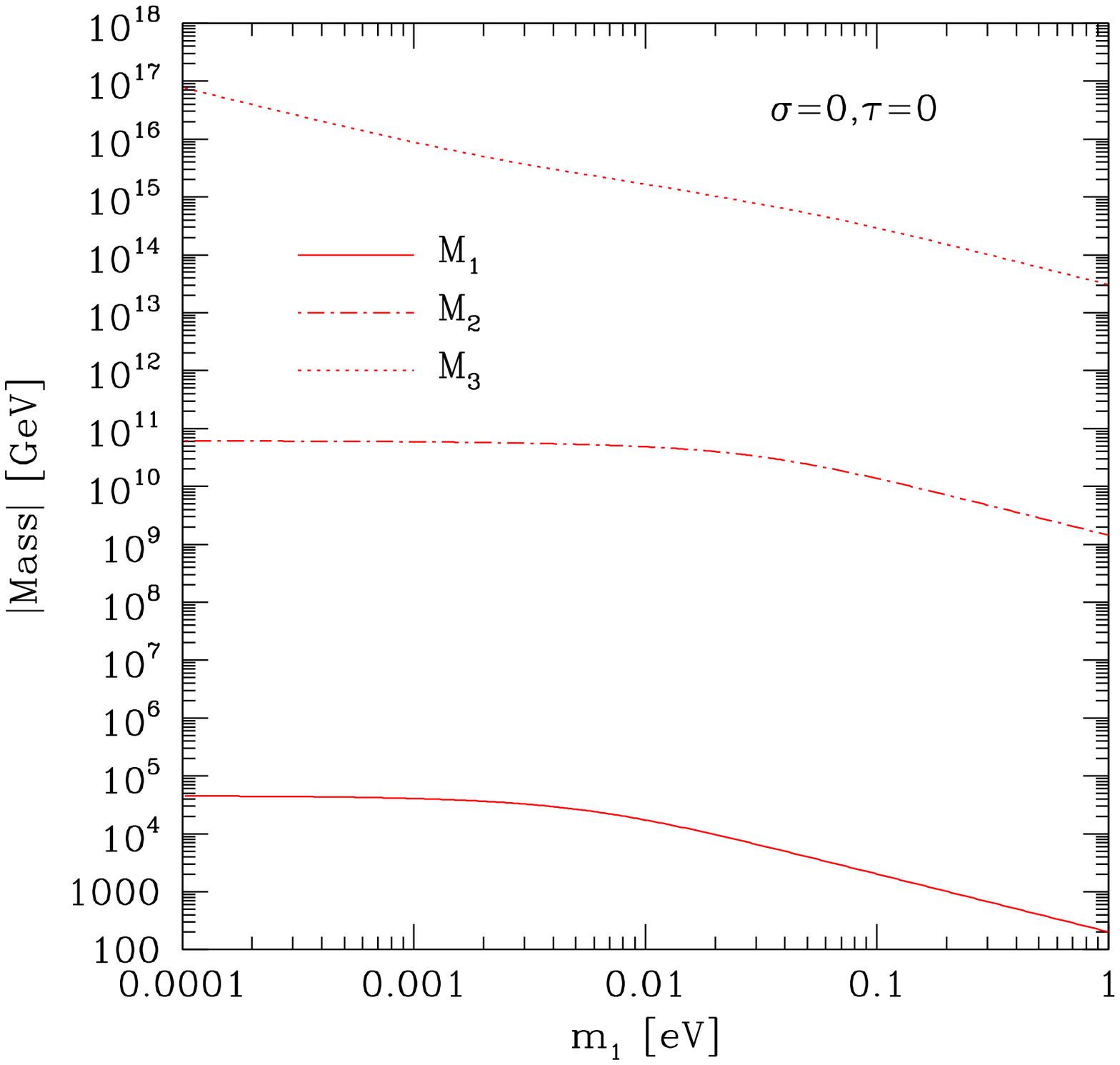,width=8.5cm,height=9cm} & 
\epsfig{file=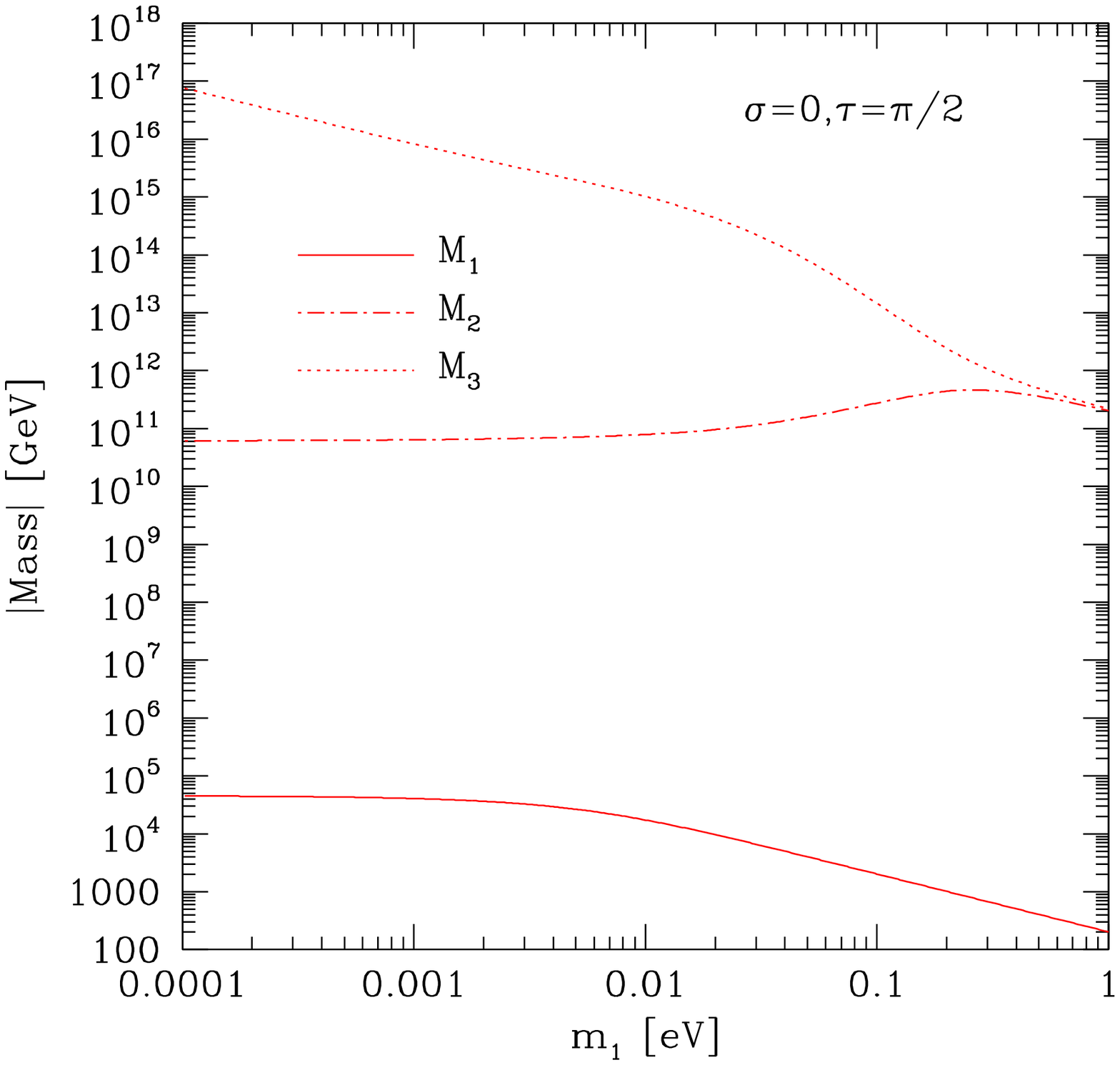,width=8.5cm,height=9cm} \\ 
\epsfig{file=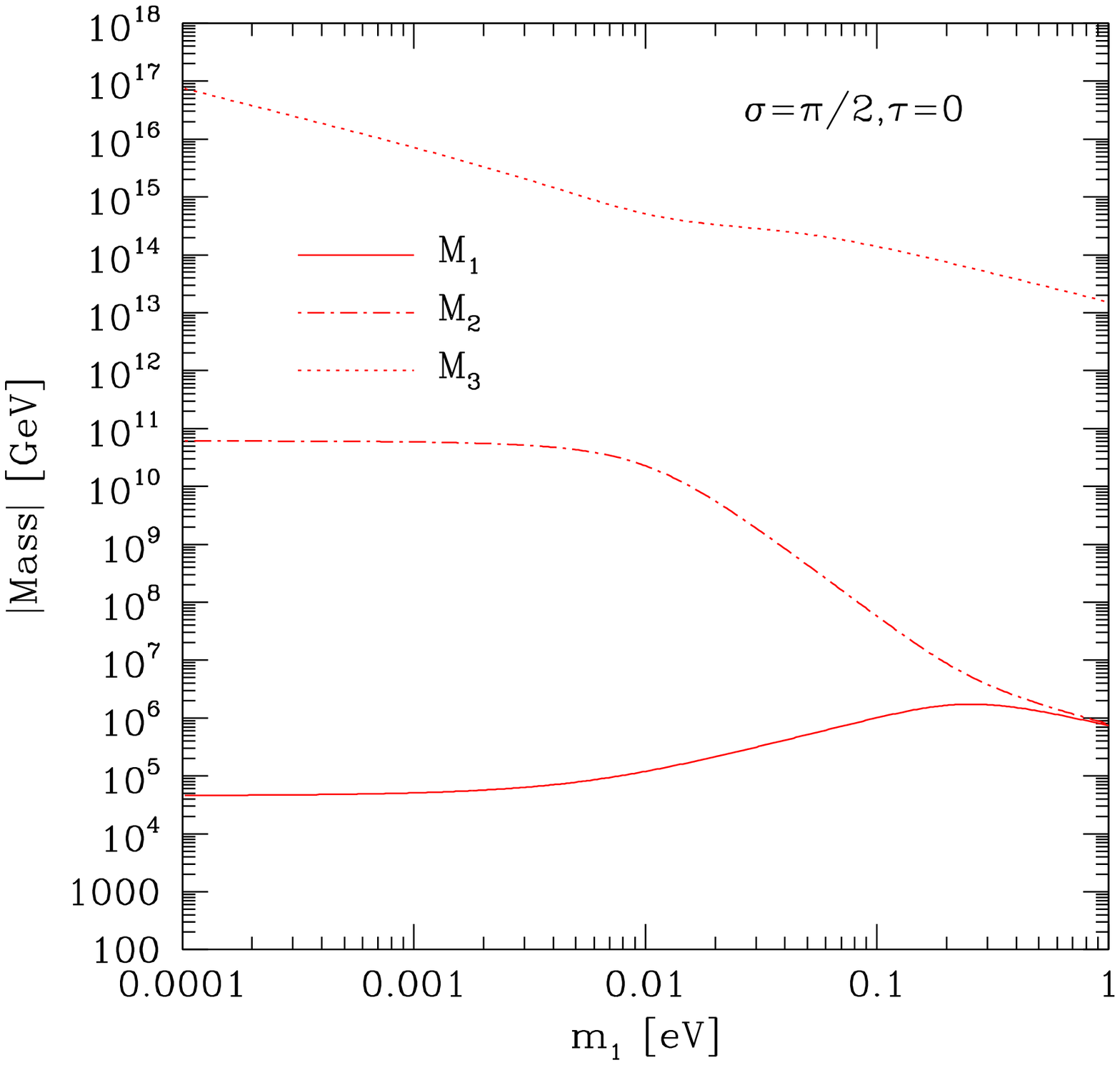,width=8.5cm,height=9cm} & 
\epsfig{file=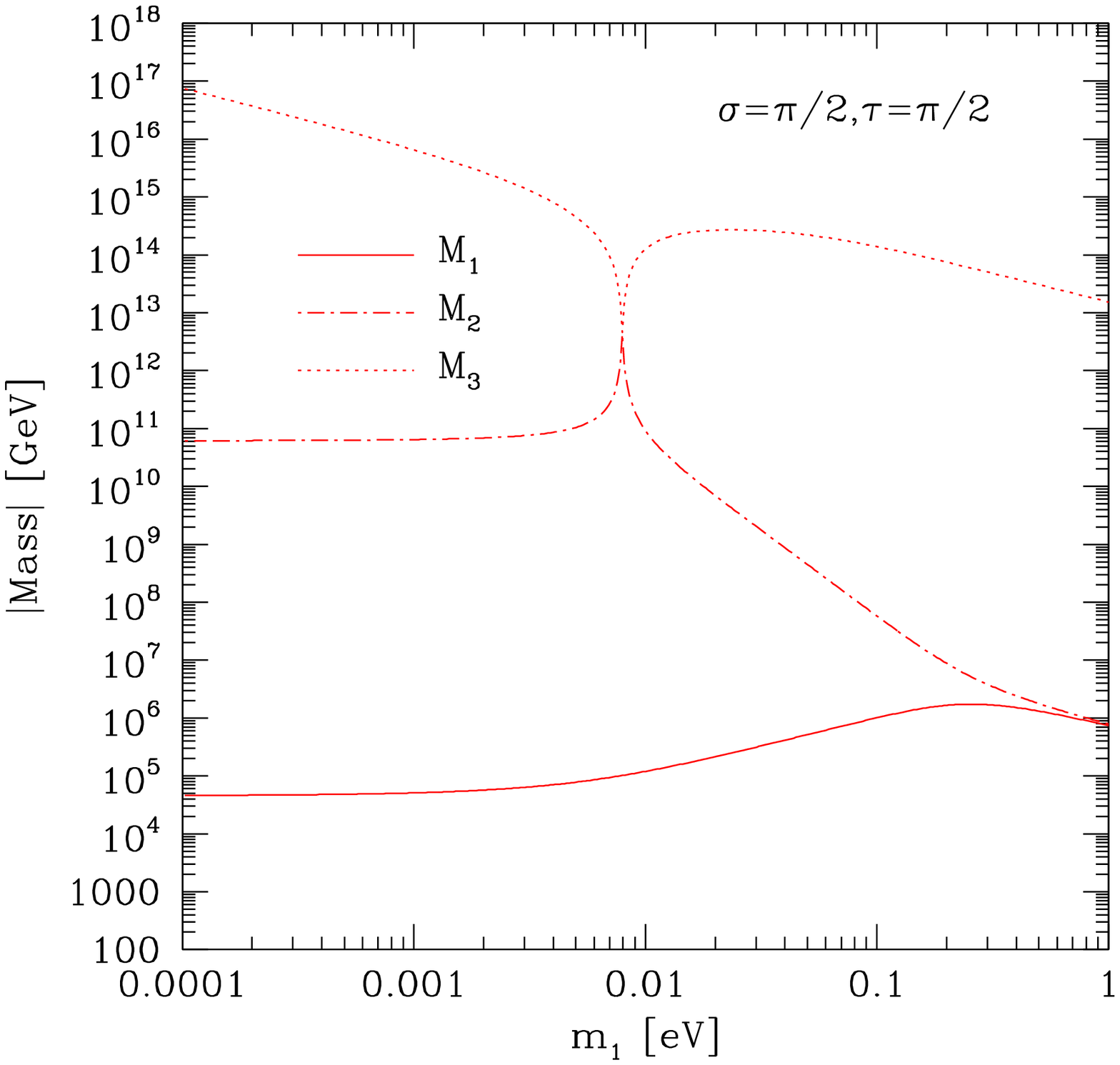,width=8.5cm,height=9cm}
\end{tabular}
\caption{\label{fig:N1}
First realization of QLC: 
the heavy neutrino masses resulting from the 
diagonalization of Eq.~(\ref{eq:MRbimax})
as a function of the smallest neutrino mass for the normal mass ordering. 
We have chosen four different pairs of values for 
$\sigma$ and $\tau$, showing the possible degeneracy of the masses. 
See text for further discussion. }
\end{figure}

\begin{figure}[htb]\hspace{-1cm}
\begin{tabular}{cc}
\epsfig{file=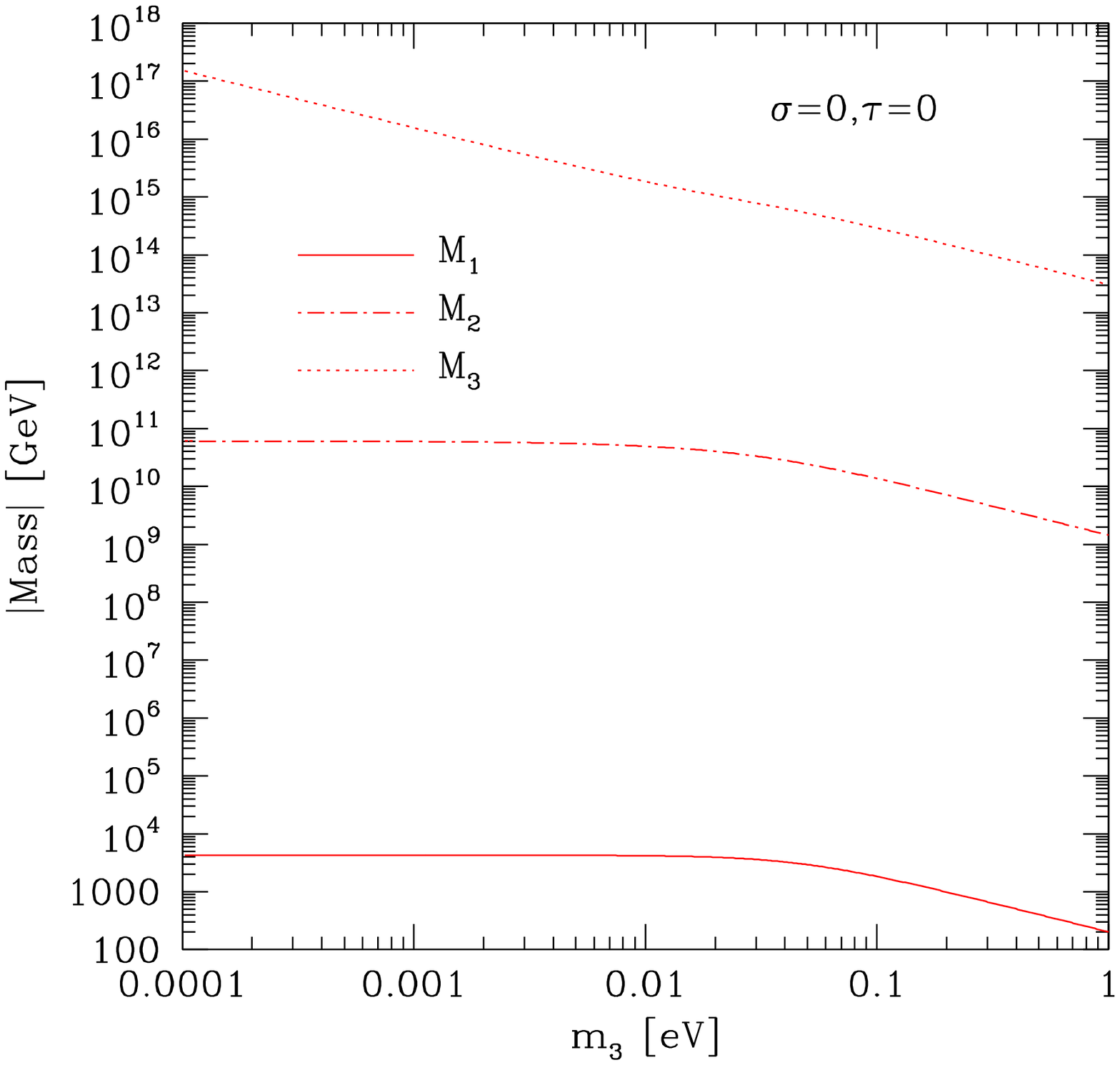,width=8.5cm,height=9cm} & 
\epsfig{file=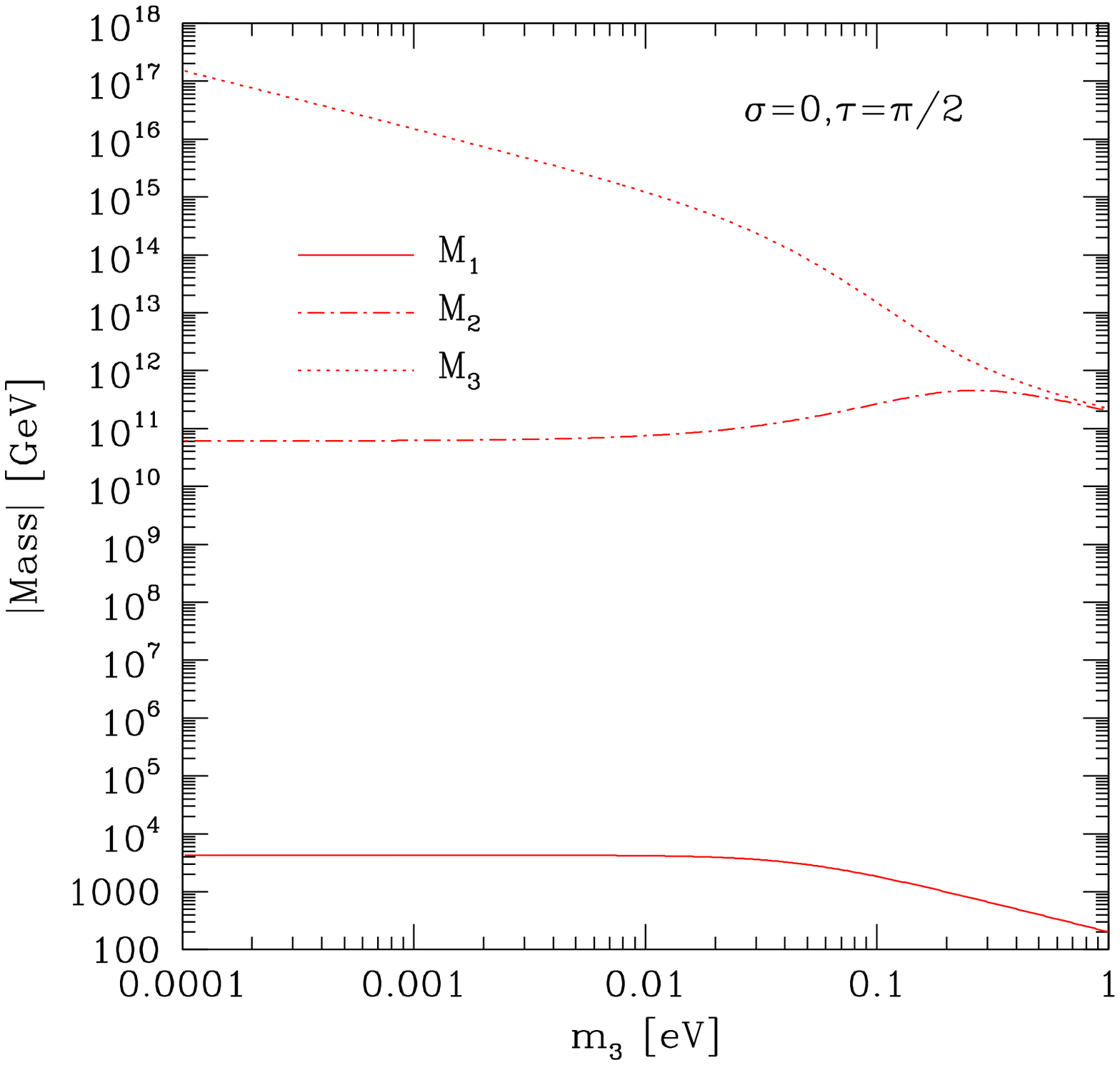,width=8.5cm,height=9cm} \\ 
\epsfig{file=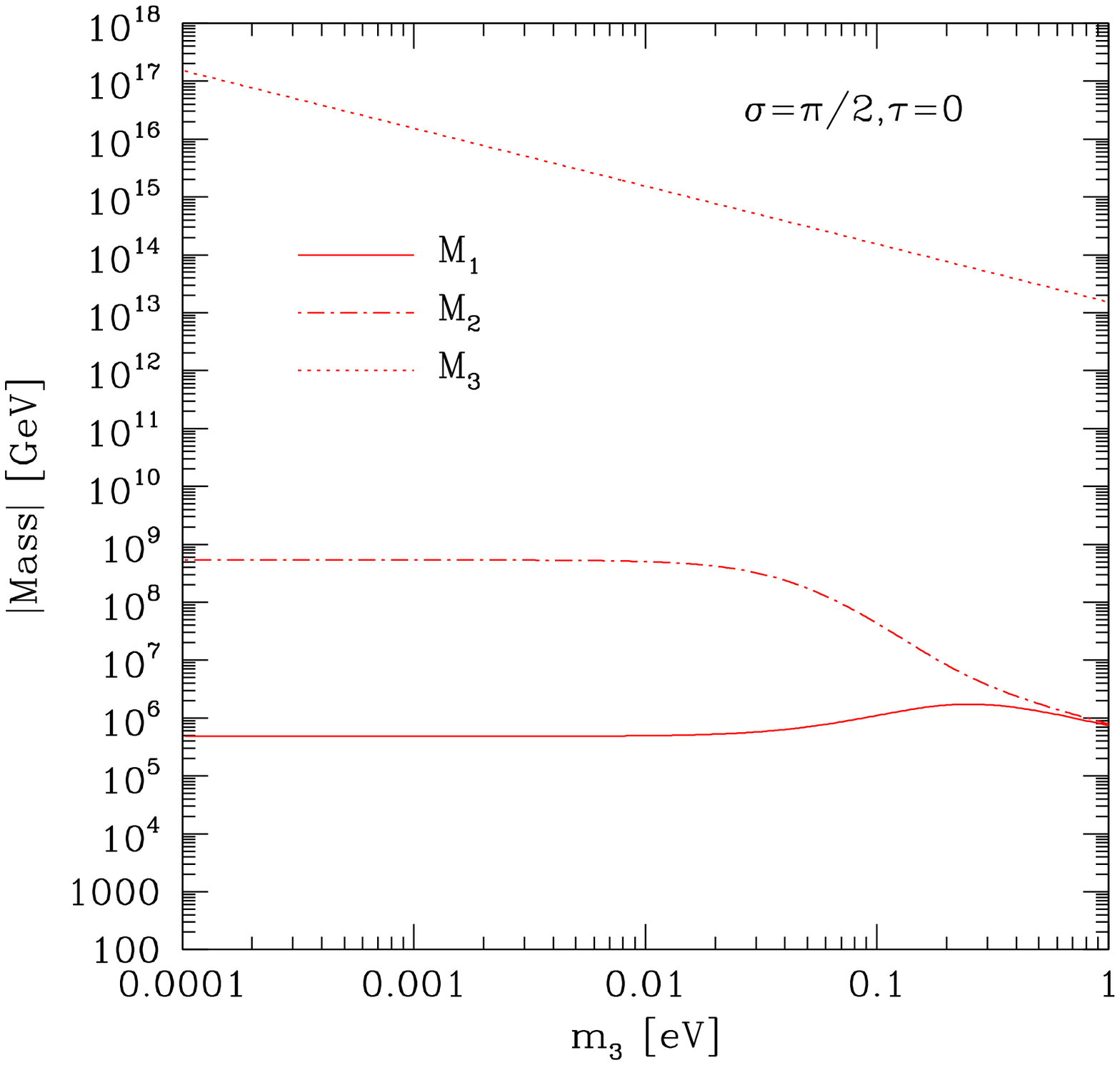,width=8.5cm,height=9cm} & 
\epsfig{file=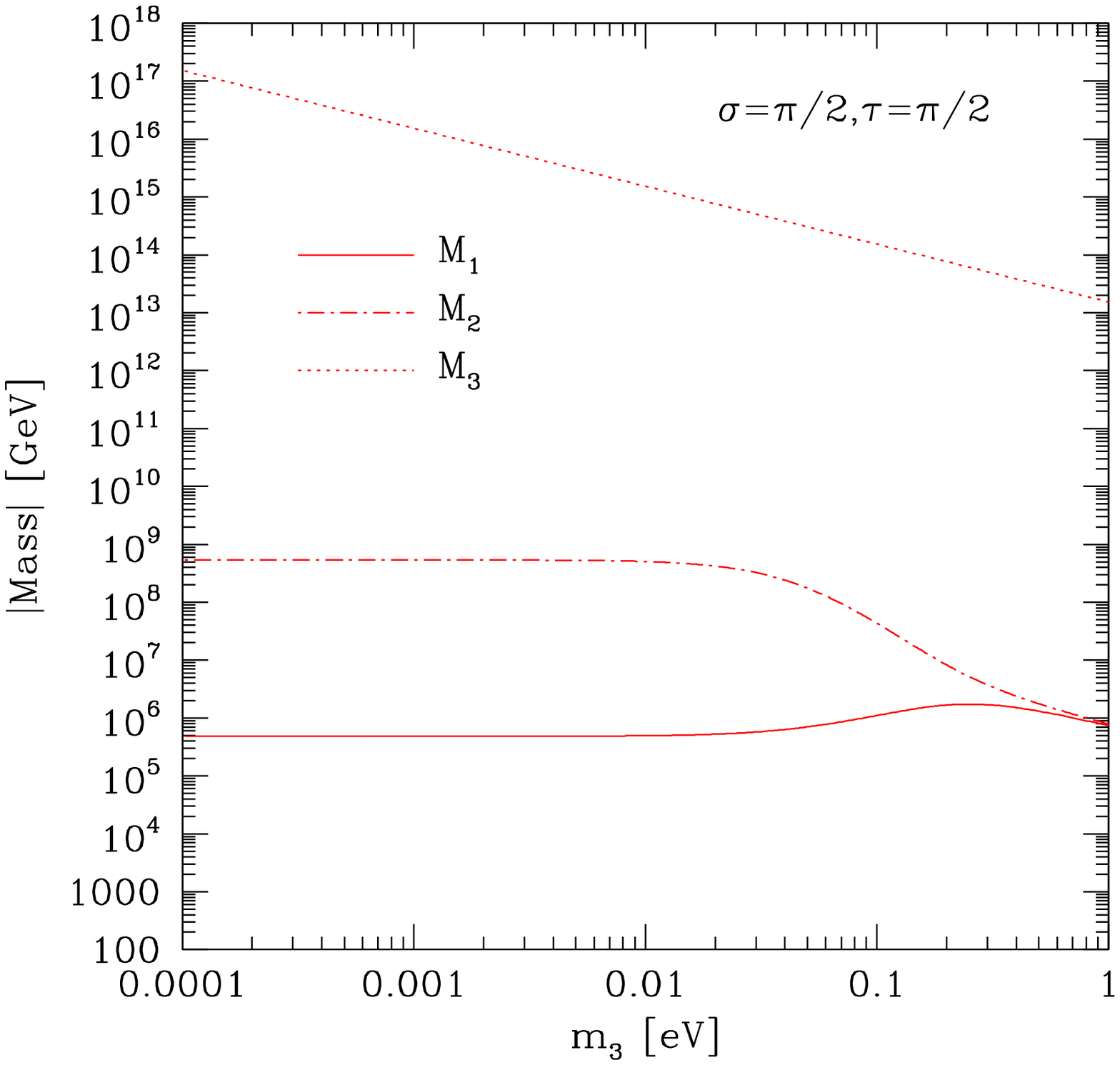,width=8.5cm,height=9cm}
\end{tabular}
\caption{\label{fig:N3}Same as Figure \ref{fig:N1} for 
the inverted mass ordering.}
\end{figure}

It is worth to discuss leptogenesis in the scenario under study. As indicated 
in Sec.~\ref{sec:YB0}, the value of the baryon asymmetry 
crucially depends on the spectrum of the heavy Majorana 
neutrinos, which we have displayed in Figs.\ \ref{fig:N1} and 
\ref{fig:N3} for normally and inversely ordered light neutrino masses. 
It also depends on the matrix $V_R$, which in case of 
$\sigma$ far away from $\pi/2$ is given in Eq.~(\ref{eq:VR}). 
In this case the eigenvalues $M_{1,2,3}$ are strongly hierarchical. In general 
$M_1$ does not exceed 
$10^{6}$ GeV, as obvious from Eq.~(\ref{eq:N1}) and 
Figs.~\ref{fig:N1}, \ref{fig:N3}.  
According to Eq.~(\ref{maxtilm1}) this is too small a 
value for successful thermal leptogenesis generated by this heavy neutrino. 
As pointed out in Section \ref{sec:YB0}, 
it is in principle possible that the second heaviest neutrino generates 
the decay asymmetry. We will illustrate now that within the QLC scenario 
under study this is problematic.   
Taking advantage of the analysis in \cite{flav1} we can estimate 
the resulting baryon asymmetry including flavor 
effects \cite{flav0,flav1}\footnote{For an analysis without 
flavor effects, see \cite{dB}.}. 
The decay asymmetry of 
the neutrino with mass $M_2$ in the flavor $\alpha = e, \mu, \tau$ 
reads \cite{flav1,dB}
\bea \label{eq:eps2}
\varepsilon_2^\alpha \D 
\simeq - \frac{1}{8 \pi\, v_u^2} 
\, \frac{1}{\left(\tilde{m}_D \, \tilde{m}_D^\dagger\right)_{22}} 
\left[ 
\frac 32 \, \frac{M_2}{M_3} \, {\rm Im} 
\left\{ (\tilde{m}_D)_{2 \alpha} \, (\tilde{m}_D^\dagger)_{\alpha 3} \, 
\left(\tilde{m}_D \, \tilde{m}_D^\dagger \right)_{23} \right\} 
\right. \\[0.2cm]
\left. \D 
+ \frac{M_1}{M_2} \, \left(\ln \frac{M_2}{M_1} - 2 \right) \, 
{\rm Im} 
\left\{ (\tilde{m}_D)_{2 \alpha} \, (\tilde{m}_D^\dagger)_{\alpha 1} \, 
\left(\tilde{m}_D \, \tilde{m}_D^\dagger \right)_{21} \right\}
\right] ~,
\eea
where $\tilde{m}_D$ is given in Eq.~(\ref{eq:mdt1}). 
In case of a normal hierarchy, we can neglect $m_1$ with respect to 
$m_2 \simeq \sqrt{\dms}$ and $m_3\simeq \sqrt{\dma}$ 
and find from Eq.~(\ref{eq:N1}) that 
\bea \label{eq:N12}
\D M_1 \, e^{i \phi_1} \simeq \frac{2 \, m_u^2}{ \sqrt{\dms}} 
\, e^{2 i \sigma} 
~~,~~
M_2 \, e^{i \phi_2} \simeq \frac{2 \, m_c^2}{\sqrt{\dma}} 
\,  e^{2 i \tau}  
~~,~~
M_3\, e^{i \phi_3} \simeq  \frac{m_t^2}{4 \, m_1}~,
\eea
which fixes $\phi_1 = 2 \sigma$, $\phi_2 = 2 \tau$ and $\phi_3 = 0$ in 
the phase matrix $R_\nu$ appearing in Eq.~(\ref{eq:VR}). 
The matrix $\tilde{V}_R$ in Eq.~(\ref{eq:VR}) simplifies considerably to 
\be
\tilde{V}_R \simeq 
\left( 
\bad \D 
1 & \D  -\frac{m_u}{\sqrt{2} \,m_c}  & \D 
\sqrt{2} \, \frac{m_u}{m_t}  \\[0.3cm]
\D  \frac{m_u}{\sqrt{2} \,m_c} & \D 1 
& \D - \frac{m_c}{m_t}  \\[0.3cm]
\D -\frac{m_u}{\sqrt{2} \,m_t}  
& \D \frac{m_c}{m_t} & 1
\ea
\right)~.
\ee
We can evaluate the decay asymmetries by making an expansion in 
terms of $\lambda$, for which we use that 
$m_c = c_c \, m_t \, \lambda^4$ and $m_u = c_u \, m_t \, \lambda^8$ 
with $c_{u,c} \simeq 1$. 
One finds that $\varepsilon_2^\tau$ is larger than 
$\varepsilon_2^{\mu}$ ($\varepsilon_2^e$) by two (four) 
orders in $\lambda$. The leading term in $\varepsilon_2^\tau$ 
comes from the contribution proportional to 
$M_2/M_3 \simeq 8 \, c_c^2 \, \lambda^8 \, m_1/m_3$ in 
Eq.~(\ref{eq:eps2}). Thus we obtain  
\be \label{eq:eps2fin}
\varepsilon_2^\tau \simeq \frac{3 \, c_c^2}{2 \, \pi \, (1 + c_c^2)} \, 
\frac{m_1}{\sqrt{\dma}} \, \lambda^8 \, \sin 2 (\omega - \phi + \tau)
\simeq 5 \cdot 10^{-9} \, \left(\frac{m_1}{10^{-4} \, \rm eV} \right) \, 
\sin 2 (\omega - \phi + \tau)~.
\ee
The second contribution  in 
Eq.~(\ref{eq:eps2}) proportional to $M_1/M_2$ is suppressed 
by $(1 - c_c^2) \, \lambda^{24} \, m_3/m_2$, which is always much smaller 
than $m_1/\sqrt{\dma} \, \lambda^8 $ due to the lower limit 
on $m_1$ from Eq.~(\ref{eq:m11}). 
We can identify the leptogenesis phase $\omega - \phi + \tau$. 
This combination of phases is not directly measurable in low energy 
experiments, as is clear from the results in Section \ref{sec:low}. 
Recall however that $\sin^2 \theta_{23} - \frac 12 
\propto \lambda^2 \, \cos(\omega - \phi)$, $J_{CP} \propto \sin \phi$ 
and $\sin \beta \simeq - \sin (\phi + \tau)$, which {\it in 
principle} allows to reconstruct the leptogenesis phase with 
low energy measurements. However, determining the Majorana phases in case 
of a normal hierarchy seems at present impossible. 
We still have to estimate the final 
baryon asymmetry from the decay asymmetry Eq.~(\ref{eq:eps2fin}). 
The wash-out of $\varepsilon_2^\tau$ by the lightest neutrino 
is governed by $\tilde{m}_1^\tau /m^\ast$, where 
$m^\ast \simeq 10^{-3}$ eV and 
$\tilde{m}_i^\tau \simeq (\tilde{m}_D)_{i \tau} 
\, (\tilde{m}_D^\dagger)_{\tau i}/M_i$. In our case, 
$\tilde{m}_1^\tau /m^\ast \simeq \sqrt{\dms}/(4 \, c_u^2 \, m^\ast) 
\simeq 2$, which confirms the result in Ref.~\cite{flav1}, where 
it has been shown that the resulting wash-out is of order 0.2. 
Without flavor effects, the wash-out would be two orders of magnitude 
stronger \cite{flav1}, which clearly demonstrates their importance. 
However, there is very strong wash-out from 
interactions involving $M_2$: the efficiency is 
$m^\ast/\tilde{m}_2^\tau \simeq 2 \, c_c^2 \, m^\ast /\sqrt{\dma}  
\simeq 1/25$ and 
the estimate for the total baryon asymmetry is \cite{flav1}
\be
\eta_B \simeq 6 \cdot 10^{-5} \, \epsilon_2^\tau 
\simeq 3 \cdot 10^{-13} \, 
\left(\frac{m_1}{10^{-4} ~\rm eV} \right) \, 
\sin 2 (\omega - \phi + \tau)~,
\ee
which is much below\footnote{This is in agreement with the 
findings of Ref.~\cite{stephane}.} the observed value of 
$6 \cdot 10^{-10}$. 
Of course, these estimates will eventually 
have to be confirmed by a precise numerical analysis. Nevertheless, 
it serves to show that successful thermal leptogenesis with the 
second heaviest Majorana neutrino is quite problematic 
in the scenario.\\

We can perform similar estimates if the 
light neutrinos are governed by an inverted 
hierarchy. After some algebra in analogy to the normal 
hierarchical case treated above we find that 
\bea \label{eq:eps2finIH}
\varepsilon_2^\tau \simeq \D \frac{3 \, c_c^2}{16 \, \pi \, (1 + c_c^2)} \, 
\frac{m_3}{\sqrt{\dma}} \, \lambda^8 \, 
\sin 2 (\phi + \tau - \omega - \sigma/2) \\ \D 
\simeq 7 \cdot 10^{-10} \, \left(\frac{m_3}{10^{-4} \, \rm eV} \right) \, 
\sin 2 (\phi + \tau - \omega - \sigma/2)~,
\eea
which is always larger than $\varepsilon_2^{e, \mu}$. This expression 
is one order of magnitude 
smaller than the decay asymmetry for the normal hierarchy. It seems therefore 
that successful leptogenesis within the inverted hierarchy is even more  
difficult. 
A more precise statement would require solving the full 
system of Boltzmann equations. 
The leptogenesis phase is now $\phi + \tau - \omega - \sigma/2$ and this 
combination of phases can in principle be reconstructed using 
$\sin^2 \theta_{23} - \frac 12 
\propto \lambda^2 \, \cos(\omega - \phi)$, $J_{CP} \propto \sin \phi$, 
$\sin \beta \simeq - \sin (\phi + \tau)$  
and $\sin (\alpha - \beta) \simeq \sin (\phi - \sigma + \tau)$. 
However, determining both Majorana phases seems at present impossible. \\

There is another interesting situation in which successful leptogenesis 
can take place in this scenario, namely 
resonant leptogenesis. 
This can occur if $\sigma \simeq \pi/2$, in which case two 
heavy neutrinos have quasi-degenerate masses, see Eq.~(\ref{eq:PD}). 
In Ref.~\cite{AFS} a similar framework was considered, and 
the mass splitting required to generate an
$\eta_B$ of the observed size has been estimated. The result corresponds 
to $|1 - M_2/M_1| \simeq 10^{-5} \ldots 10^{-6}$, 
which is a rather fine-tuned situation. 
However, there are two rather interesting aspects to this case: 
as discussed in Section \ref{sec:low}
the phase $\sigma$ is related to the low energy Majorana phase $\alpha$. 
If $\alpha = \pi/2$ it is known that for quasi-degenerate 
neutrinos the stability with respect to radiative corrections 
is significantly improved \cite{chef}. Moreover, 
the resonant condition occurs if the smallest neutrino mass 
is approximately 0.5 eV, i.e., the light neutrinos are quasi-degenerate.
In this case the effective mass for \onbb reads 
\be
\meff \simeq m_1 \, \left( \sqrt{2} \, \lambda + \frac 12 \, 
c_{\phi + 2 \tau} \, \lambda^2 
\right)~.  
\ee
The maximum value of the effective mass for quasi-degenerate neutrinos 
is roughly $m_1$ \cite{0vbbrev}. The suppression factor 
$\sqrt{2} \, \lambda$ is nothing but $\cos 2 \theta_{12}$. 
Therefore there are sizable cancellations in the effective mass 
\cite{ichNPB} when 
the resonance condition for the heavy neutrino masses is fulfilled. 
With $m_1 \simeq 0.5$ eV we can predict that 
$\meff \simeq 0.16$ eV, a value which can be easily tested in running 
and up-coming experiments \cite{0vbbAPS}.\\

If $\tau \simeq \pi/2$, it is apparent from  
Figs.~\ref{fig:N1} and \ref{fig:N3} that situations can occur 
in which $M_2$ and $M_3$ are quasi-degenerate. Hence, their decay 
could create a 
resonantly enhanced decay asymmetry, but the lighter 
neutrino with mass $M_1$ should not wash out this asymmetry. 
Determining if this is indeed possible would again require a dedicated 
study and solution of the Boltzmann equations. 
Leaving the fine-tuned possibility of resonant leptogenesis aside, 
we can consider non-thermal leptogenesis. 
However, as also discussed in Ref.~\cite{AFS}, 
the decay asymmetry $\varepsilon_1$ turns out to be too tiny: 
if we insert the phenomenological 
values $m_u/m_c \sim m_c/m_t \simeq \lambda^4$ in the 
exact equations and if we refrain from considering the 
possibility of resonant enhancements, $\varepsilon_1$ is of 
order $\lambda^{16} \simeq 10^{-11}$. In principle, 
the baryon asymmetry could be generated by the decays of the 
heavier neutrinos, i.e., by $\varepsilon_2$ and/or 
$\varepsilon_3$, which are indeed larger than $\varepsilon_1$. 
This possibility would indicate that the inflaton has a sizable 
branching ratio in the heavier neutrinos.  
However, this would also require that the lightest Majorana neutrino  
$N_1$ does not wash out the asymmetry generated by $N_2$ and $N_3$, making 
a detailed numerical analysis necessary.

\section{\label{sec:second}Second Realization of QLC}
In this section we discuss another possible 
realization of QLC, which has also 
been outlined already in \cite{QLC0,QLC1}: 
\begin{itemize}
\item the conventional see-saw mechanism generates the neutrino mass matrix. 
Diagonalization of $m_\nu$ is achieved via 
$m_\nu = U_\nu^\ast \, m_\nu^{\rm diag} \, U_\nu^\dagger$ and $U_\nu$ is 
related to $V$ (in the sense that 
$U_\nu^\dagger = P_\nu \, V \, Q_\nu$); 
\item the matrix diagonalizing the charged leptons corresponds 
to bimaximal mixing: $U_\ell =  U_{\rm bimax}^T$. This can be achieved 
when $V_{\rm up} = V^\dagger$, therefore 
$V_{\rm down} = \mathbbm{1}$;  
\item if indeed $V_{\rm up} = V^\dagger$, then  
$m_\nu = -m_D^T \, M_R^{-1} \, m_D$ is diagonalized by the CKM matrix. 
If $M_R$ does not introduce additional rotations we can have the 
$SO(10)$-like relation 
$m_{\rm up} = m_D = V'_{\rm up} \, m_{\rm up}^{\rm diag} 
\, P_\nu \,  V \, Q_\nu$. Here $V'_{\rm up}$ denotes 
in our convention the in principle unknown right-handed rotation of $m_D$. 
The condition of $M_R$ not introducing additional rotations 
means that $V_R = (V'_{\rm up})^\ast$, where 
$M_R = V_R^\ast \, M_R^{\rm diag} \, V_R^\dagger$. 

\end{itemize}

Note that 
the equalities  $U_\ell =  U_{\rm bimax}^T$ and $V_{\rm down} = \mathbbm{1}$ 
are consistent with the $SU(5)$-like relation 
$m_\ell = m_{\rm down}^T$. The same comments 
as in the first realization of QLC on 
whether the indicated scenario is ``realistic'' or not, 
would then apply here. If $m_\ell = m_{\rm down}^T$ was not assumed, 
the quark and lepton sector would not be related.\\ 

\begin{figure}[htb]\hspace{-1cm}
\epsfig{file=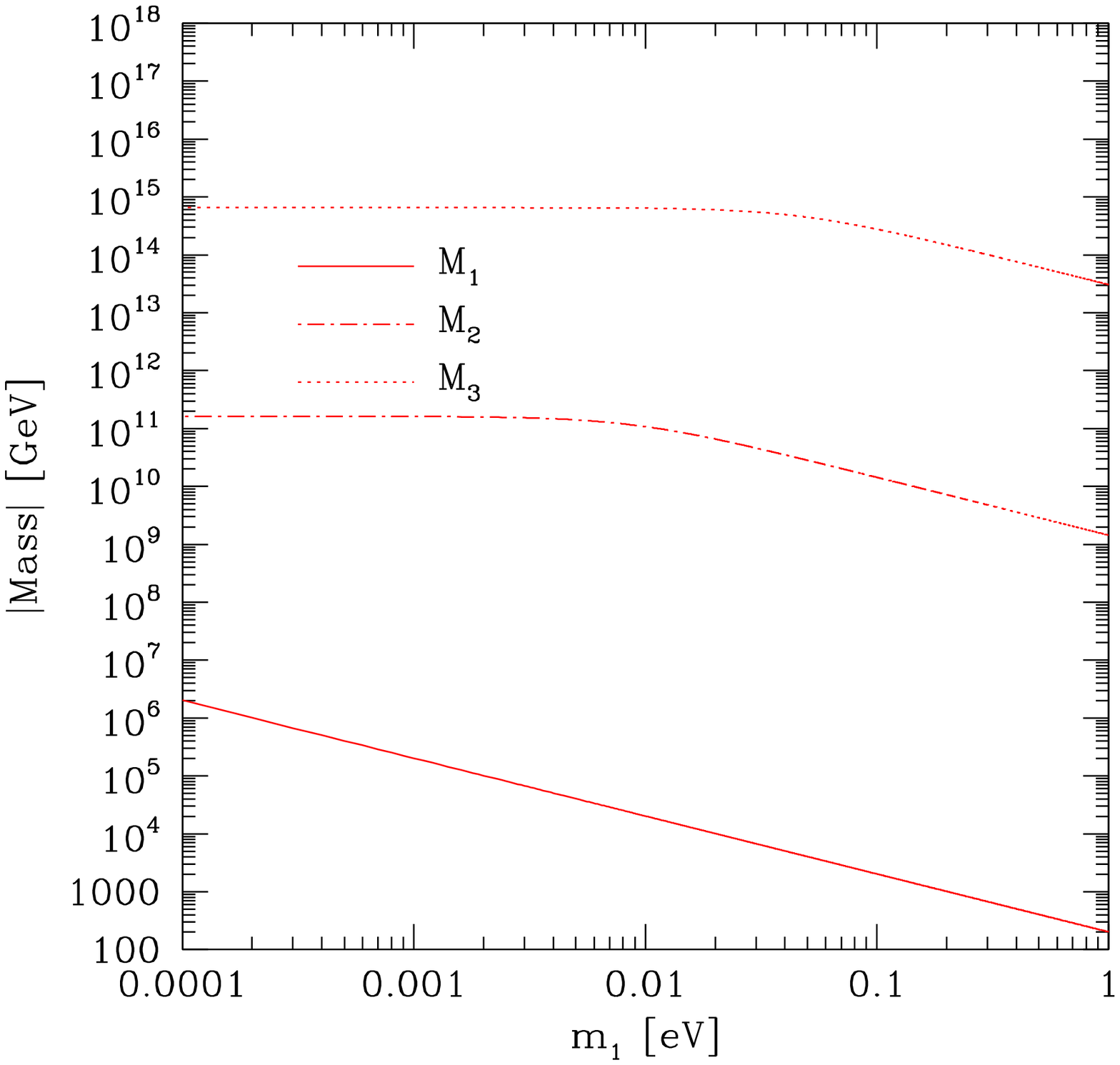,width=8.9cm,height=9cm} 
\epsfig{file=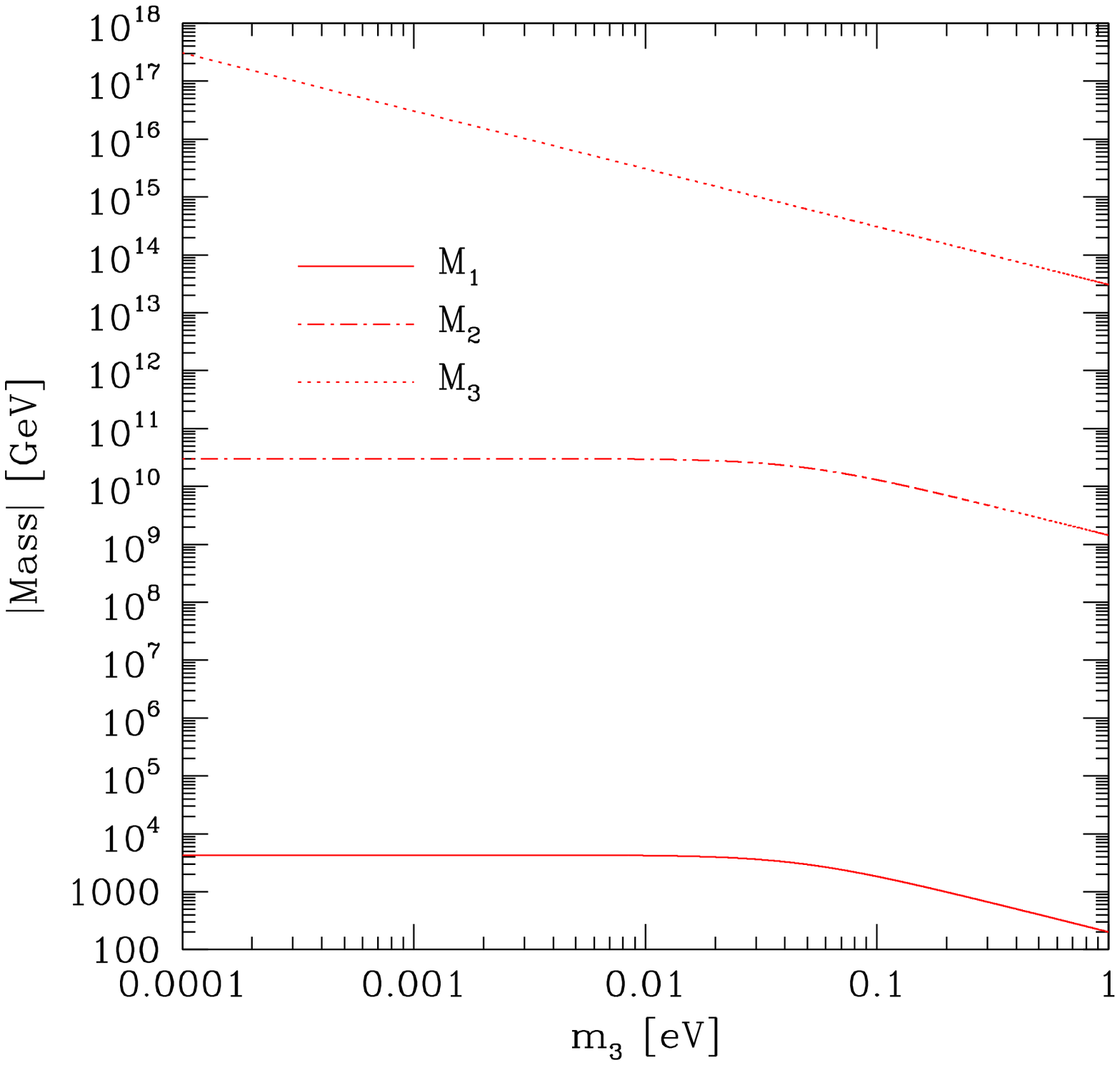,width=8.9cm,height=9cm} 
\caption{\label{fig:N12}
Second realization of QLC: 
the heavy neutrino masses 
as a function of the smallest neutrino mass for the normal (left plot) 
and inverted (right plot) mass ordering.}
\end{figure}

In the following we will redo the calculations of the previous sections 
for all the observables with this second set of assumptions. 
First of all we note that in the important basis in which the 
charged leptons and heavy neutrinos are real and diagonal the 
Dirac mass matrix reads 
\bea \label{eq:mdt2}
m_D \rightarrow 
\tilde{m}_D = V_R^T \, m_D \, U_\ell 
= m_{\rm up}^{\rm diag} \, P_\nu \, V \, Q_\nu \, 
U_{\rm bimax}^T \\[0.2cm] 
\Rightarrow 
\left\{ 
\baz 
\tilde{m}_D^\dagger \, \tilde{m}_D 
= U_{\rm bimax} \, Q_\nu^\dagger \, V^\dagger \, 
{\rm diag}(m_u^2, m_c^2, m_t^2) \, V \, Q_\nu \, 
U_{\rm bimax}^T
& \mbox{ for LFV}~,\\[0.3cm]
\tilde{m}_D \, \tilde{m}_D^\dagger 
= {\rm diag}(m_u^2, m_c^2, m_t^2) &  \mbox{ for } \eta_B
~.
\ea 
\right. 
\eea
The correspondence between the light and heavy Majorana neutrino masses 
is rather trivial: 
\be \label{eq:masses2}
M_1 = \frac{m_u^2}{m_1}~~,~~M_2 = \frac{m_c^2}{m_2}~~,~~
M_3 = \frac{m_t^2}{m_3}~.
\ee
In Fig.\ \ref{fig:N12} we show the neutrino masses as a function 
of the smallest neutrino mass $m_1$ and $m_3$ for the normal 
and inverted ordering, respectively. Again, we have taken the 
best-fit points for \dms~and \dma~and 
the quark masses are $m_u = 0.45$ MeV, $m_c = 1.2$ GeV 
and $m_t = 175 $ GeV. Note that in contrast to the first realization of QLC 
there is no possibility to 
enhance the neutrino masses, since they 
do not depend on phases. We can set a lower limit on $m_1$ or $m_3$ which 
stems from the requirement that $M_1$ or $M_3$ 
does not exceed the Planck mass: 
\be \label{eq:m12}
m_1 \ge \frac{m_u^2}{M_{\rm Pl}} \simeq 8 \cdot 10^{-17}~{\rm eV} 
\mbox{ and } 
m_3 \ge \frac{m_t^2}{M_{\rm Pl}} \simeq 2 \cdot 10^{-7}~{\rm eV} ~.
\ee
This is for $m_1$ 10 orders of magnitude smaller than the corresponding limit 
in the first realization of QLC, see Eq.~(\ref{eq:m11}).\\

Interestingly, there can be no leptogenesis in this scenario. First of all, 
$M_1$ is lighter than $10^7$ GeV and this is -- in analogy to 
the first realization of QLC -- too small a value for successful leptogenesis. 
Can the decay of the second heaviest neutrino generate the baryon asymmetry? 
The answer is no, simply because 
$\tilde{m}_D \, \tilde{m}_D^\dagger$ is diagonal, as can be seen in 
Eq.~(\ref{eq:mdt2}). 
The decay asymmetries, both in the case when one 
sums over all flavors, Eq.~(\ref{e1H}), and the asymmetries 
for a given flavor, Eq.~(\ref{eq:eps2}), are always proportional to 
off-diagonal entries of $\tilde{m}_D \, \tilde{m}_D^\dagger$ and therefore 
always vanish in this realization of QLC.

\subsection{\label{sec:low2}Low Energy Neutrino Phenomenology}
In our second case the PMNS matrix can be written as 
\be \label{eq:def2}
U = U_\ell^\dagger \, U_\nu = R_{23} \, R_{12} \, 
(P_\nu \, V \, Q_\nu)^\dagger ~, 
\ee
where $R_{ij}$ is a rotation 
with $\pi/4$ around the ($ij$)-axis and $P_\nu$ and $Q_\nu$ are defined in 
Eq.~(\ref{eq:Ubima}). 
We remark that an analysis of this framework 
including all possible phases has not been performed before 
(see Refs.~\cite{QLC0,QLC1,2others}). 
With our parameterization of 
the PMNS matrix the two phases in $P_\nu$ are ``Majorana-like'' and do not 
show up in oscillations. All phases originate from the neutrino sector. 
The neutrino oscillation observables are  
\be \label{eq:QLCobs2}
\ba
\sin^2 \theta_{12} = \frac{\D 1}{\D 2} 
- \lambda \, \cos \sigma  + {\cal O}(\lambda^3) ~,\\[0.2cm]
|U_{e3}| = \frac{\D A }{\D \sqrt{2}} \, \lambda^2 
+ {\cal O}(\lambda^3) ~,\\[0.2cm]
\sin^2 \theta_{23} = \frac{\D 1}{\D 2} - 
\sqrt{\frac 12} \, A \, \lambda^2 \, 
\cos (\tau - \sigma) + {\cal O}(\lambda^3)~,\\[0.2cm]
J_{CP}^{\rm lep} = \frac{\D \lambda^2}{\D 4\sqrt{2}} \, 
\sin(\tau - \sigma) + {\cal O}(\lambda^3)~.
\ea 
\ee
The solar neutrino mixing parameter depends on the $CP$ phase $\sigma$. 
Note that in order to have solar neutrino mixing of the observed 
magnitude, the phase has to be close to zero or 
$2\pi$, at $3\sigma$ typically below $\pi/4$ 
(or above $7\pi/4$). 
The prediction for $\sin^2 \theta_{12}$ is\footnote{
Again, we do not use the approximate expressions in 
Eq.~(\ref{eq:QLCobs2}), but the exact equations. 
Besides the phases, we also vary the parameters of the 
CKM matrix in their 1, 2 and 3$\sigma$ 
ranges, and also fix these parameters to their best-fit values.} 
\be
\sin^2 \theta_{12} \gs 0.279~(0.278, \, 0.277, \, 0.276)~. 
\ee
These are lower values than in our first scenario. The numbers have 
to be compared to 
the $1\sigma$ ($2\sigma$) limit of $\sin^2 \theta_{12} \le 0.33~(0.37)$. 
The parameter $|U_{e3}|$ has a ``central value'' of 
$A\, \lambda^2/ \sqrt{2} \simeq 0.0295$. In the 
first scenario the prediction was 
$|U_{e3}|^2 = 0.0258$, which is by chance almost 
the same number. We find a range of 
\be
|U_{e3}| = 0.0295^{+0.0059, \, 0.0070, \, 0.0085, \, 0.0099}
_{-0.0058, \, 0.0066, \, 0.0076, \, 0.0084}
~.
\ee
Recall that the $1\sigma$ ($2\sigma$) bound on $|U_{e3}|$ is 0.11 (0.17). 
Probing such small values of $|U_{e3}|$ is rather challenging and 
would require at least superbeams \cite{exp_new}.
Due to cancellations $\sin^2 \theta_{23} = \frac 12$ can always 
occur. In this case, $\cos (\tau - \sigma) = 0$ and 
$J_{CP}^{\rm lep}$ takes its maximal possible value. 
The minimal and maximal values of $\sin^2 \theta_{23}$ are 
given by 
\be
\sin^2 \theta_{23} \ge 0.466~(0.465, \, 0.463, \, 0.462)~\mbox{ and } 
\sin^2 \theta_{23} \le 0.536~(0.538, \, 0.539, \, 0.540)~,
\ee
which is only a slightly larger range compared to the 
first scenario, and thus hard to probe experimentally. 
Leptonic $CP$ violation is in leading order proportional to 
$\lambda^2 \, \sin (\tau - \sigma)$, which is four powers of 
$\lambda$ larger than the $J_{CP}$ of the quark sector. 
If the neutrino sector conserved $CP$, then $J_{CP}^{\rm lep}$ vanishes. 
Note that the phase combination ($\tau - \sigma$) governs the 
magnitude of the atmospheric neutrino mixing. 
In the first scenario, $J_{CP}^{\rm lep}$ and the solar neutrino mixing 
were correlated in this way. 
In analogy to Eq.~(\ref{eq:sumrule}) we can write the sum-rule 
\bea \label{eq:sumrule2}
\sin^2  \theta_{23} \simeq \frac 12 - |U_{e3}| \, \cos (\tau - \sigma) 
\simeq  \frac 12 \pm \sqrt{|U_{e3}| - 16 \, J_{CP}^2}~.
\eea
In Fig.~\ref{fig:the_one_ring2} we show the correlations between the 
oscillation parameters which result from the relation 
in Eq.~(\ref{eq:def2}). 
We plot $J_{CP}^{\rm lep}$ and $\sin^2 \theta_{12}$ 
against $\sin^2 \theta_{23}$, as well as 
$\sigma$ and  $|U_{e3}|$ against 
$\sin^2 \theta_{12}$. We also indicate the current 
1, 2 and $3\sigma$ ranges of the oscillation parameters, showing that 
the predictions of this scenario are perfectly compatible with 
all current data.\\

Turning aside again from the oscillation observables, 
the invariants for the Majorana phases are 
\be \label{eq:S122}
S_1 
= -\frac{\D \lambda^2}{\D 2} \, A \, \sin (\sigma + \omega) 
+ {\cal O}(\lambda^3)\mbox{ and } 
S_2  
= -\frac{\D \lambda^2}{\D 2} \, A \, \sin (\omega - \phi)  
+ {\cal O}(\lambda^3)
~.
\ee
In analogy to the discussion following Eq.~(\ref{eq:S121}), we 
can translate these formulae into expressions for the 
low energy Majorana phases $\alpha$ and $\beta$. 
This leads to $\sin \beta \simeq \sin (\sigma + \omega)$ and 
$\sin (\alpha - \beta) \simeq \sin (\phi - \omega)$ and indicates that 
$\alpha$ in the parameterization of Eq.~(\ref{eq:Upara}) is related to 
($\phi + \sigma$). Indeed, a calculation of the effective 
mass in case of an inverted hierarchy, 
where the Majorana phase $\alpha$ plays a 
crucial role \cite{0vbbrev}, results in  
\be
\meff \simeq \sqrt{\dma} \, \left|  c_{\phi + \sigma} 
+ 2 \frac{s_\phi}{c_{\phi + \sigma} } \lambda^2 
\right| ~.
\ee
Similar statements can be made for quasi-degenerate neutrinos.

\subsection{\label{sec:LFV2}Lepton Flavor Violation}
\begin{figure}[htb]\hspace{-1cm}
\epsfig{file=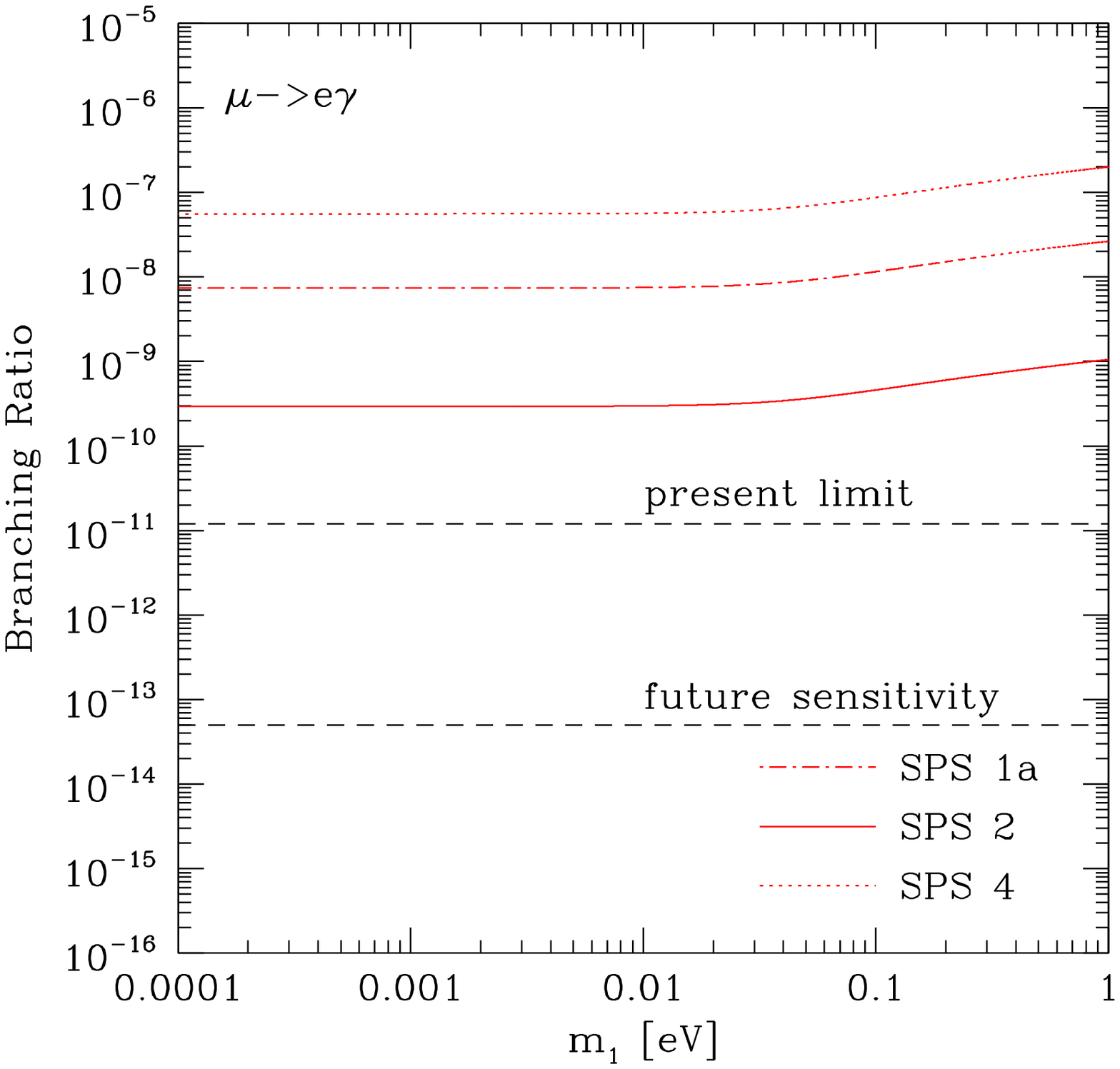,width=5.6cm,height=8cm}
\epsfig{file=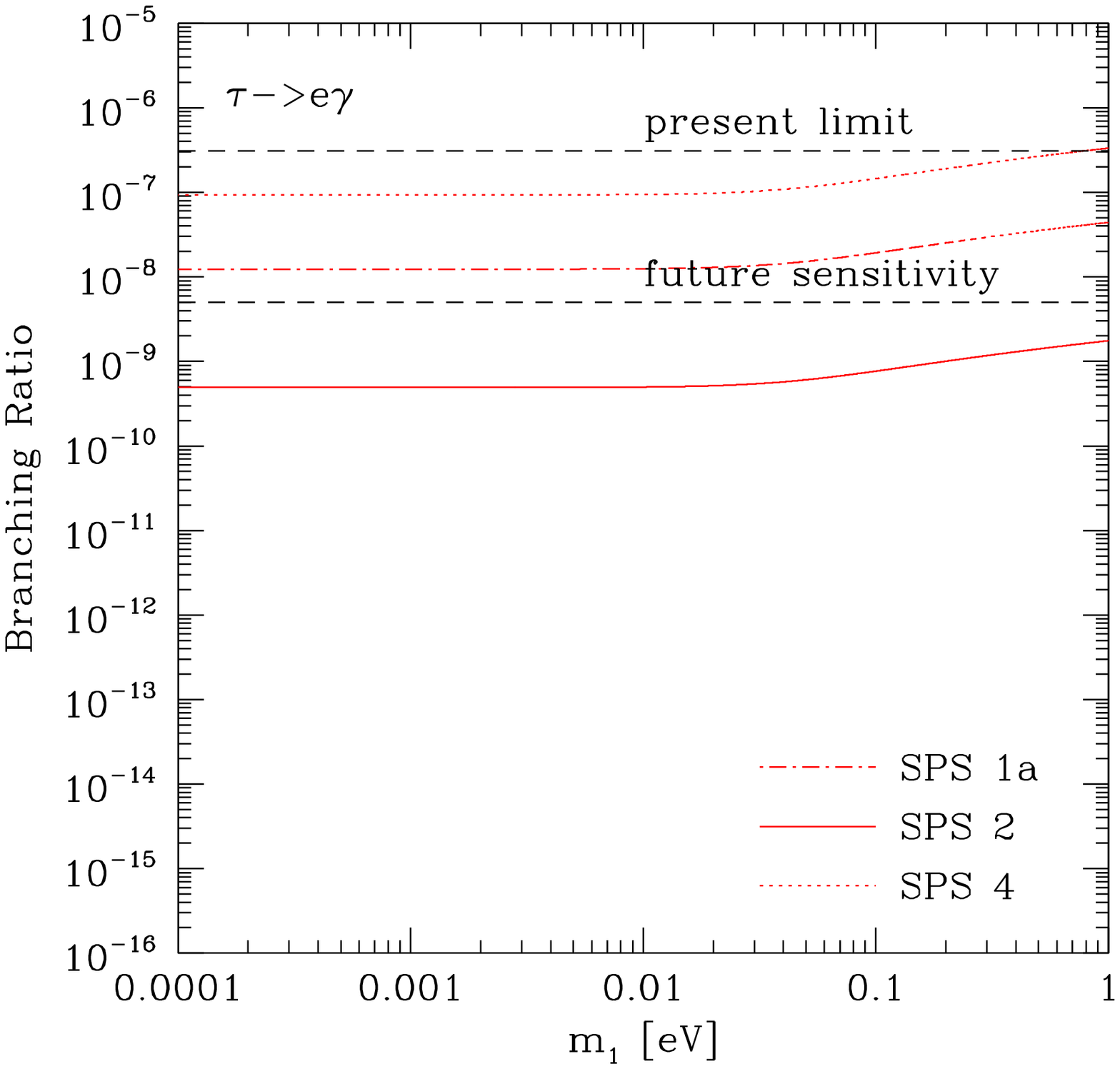,width=5.6cm,height=8cm}
\epsfig{file=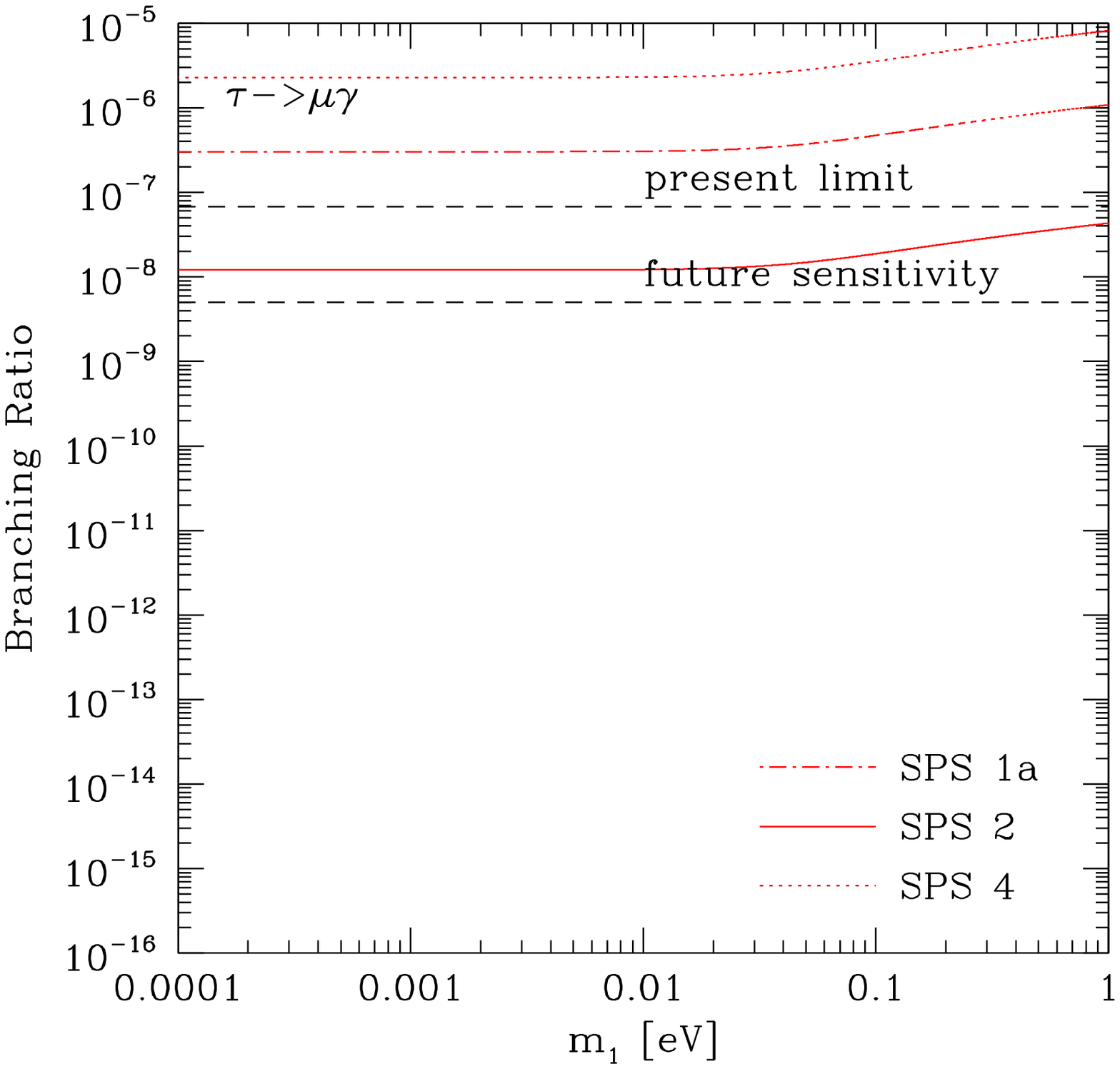,width=5.6cm,height=8cm}
\caption{\label{fig:BR2NH}Second realization of QLC: 
the branching ratios for $\mu \ra e \gamma$, 
$\tau \ra e \gamma$ and $\tau \ra \mu \gamma$ against the smallest 
neutrino mass (normal ordering) 
for the SPS points 1a, 2 and 4, see Table \ref{tab:SPS}. 
Indicated are also the present and future sensitivities.}
\end{figure}

\begin{figure}[htb]\hspace{-1cm}
\epsfig{file=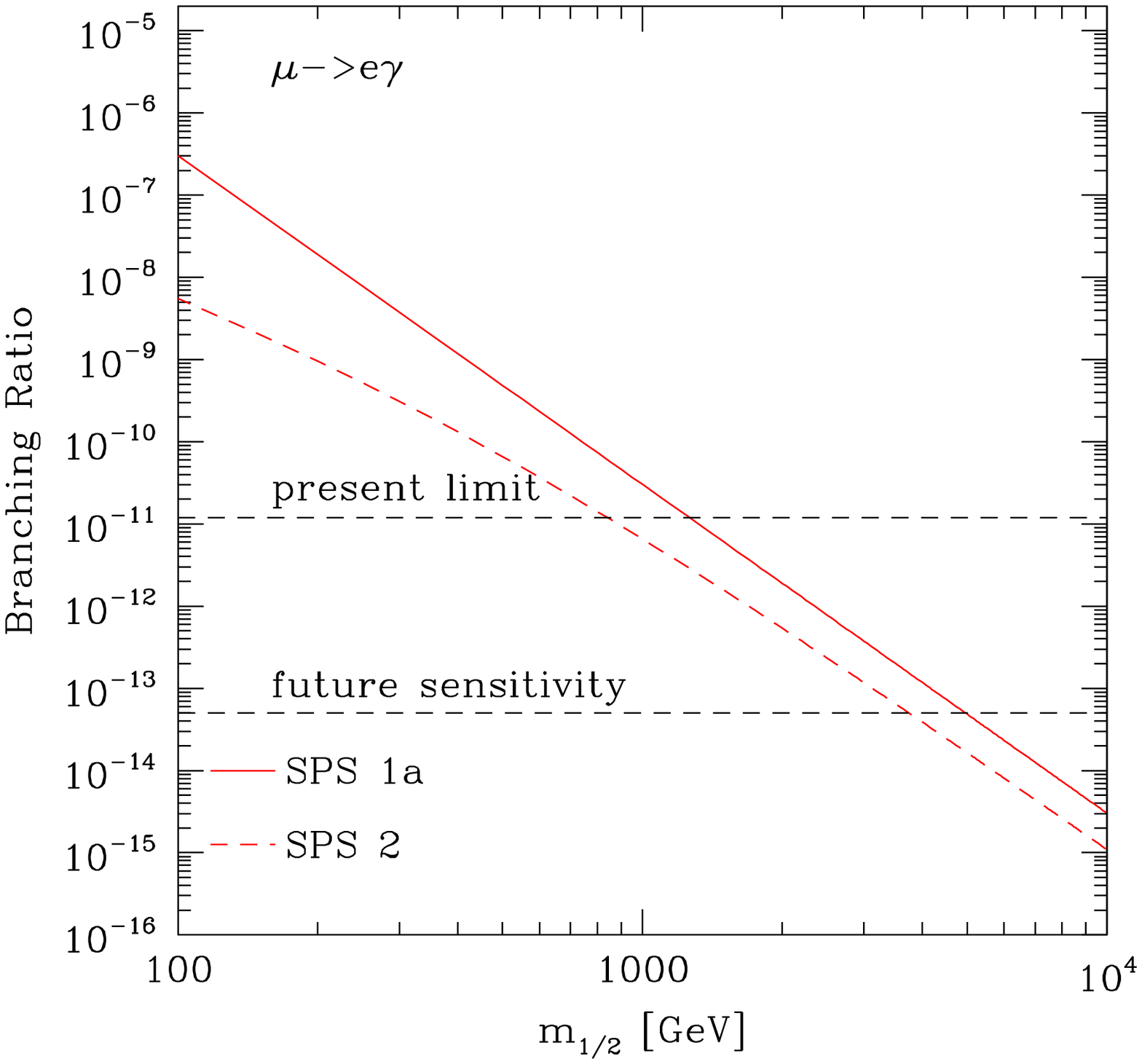,width=5.6cm,height=8cm}
\epsfig{file=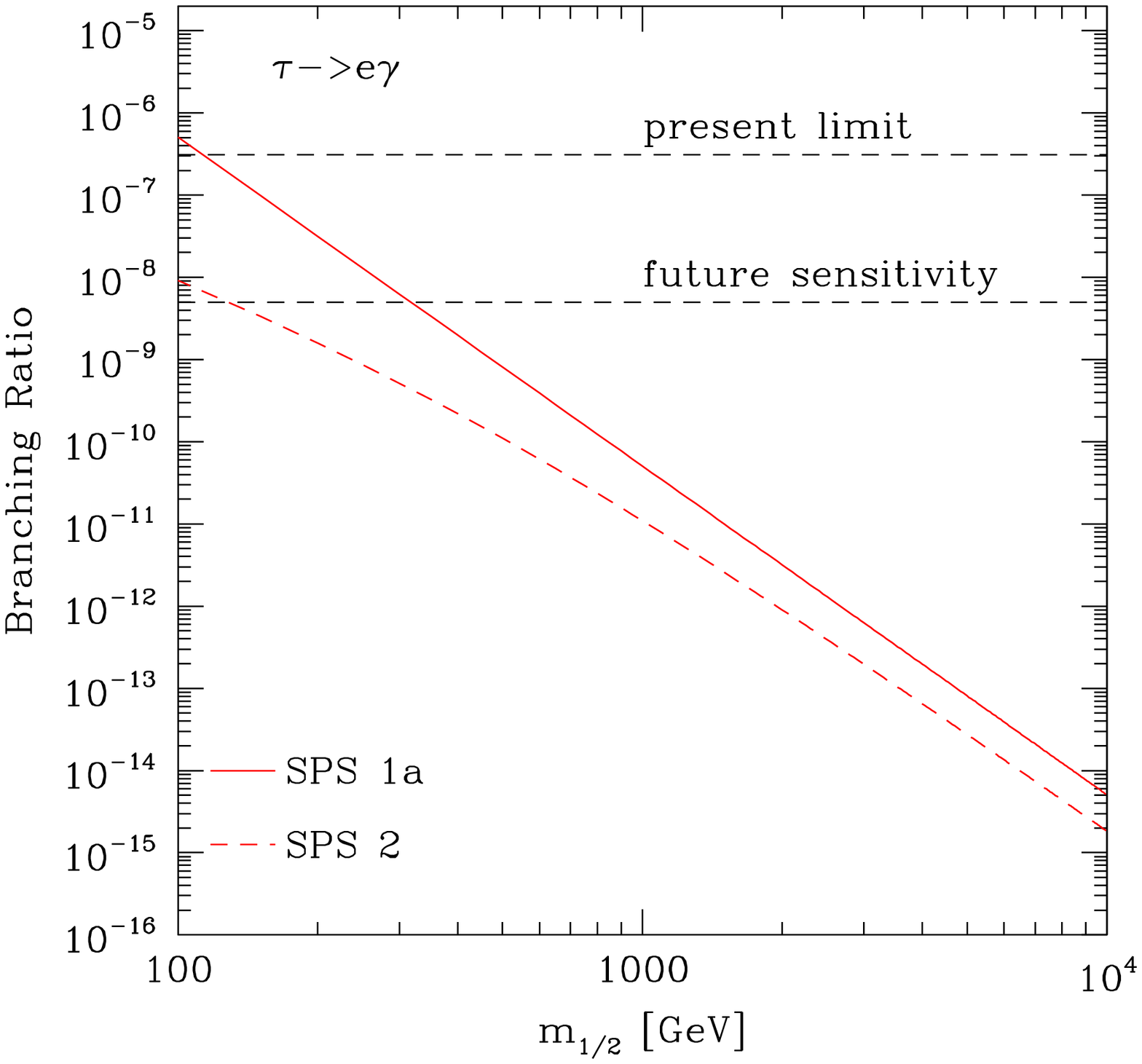,width=5.6cm,height=8cm}
\epsfig{file=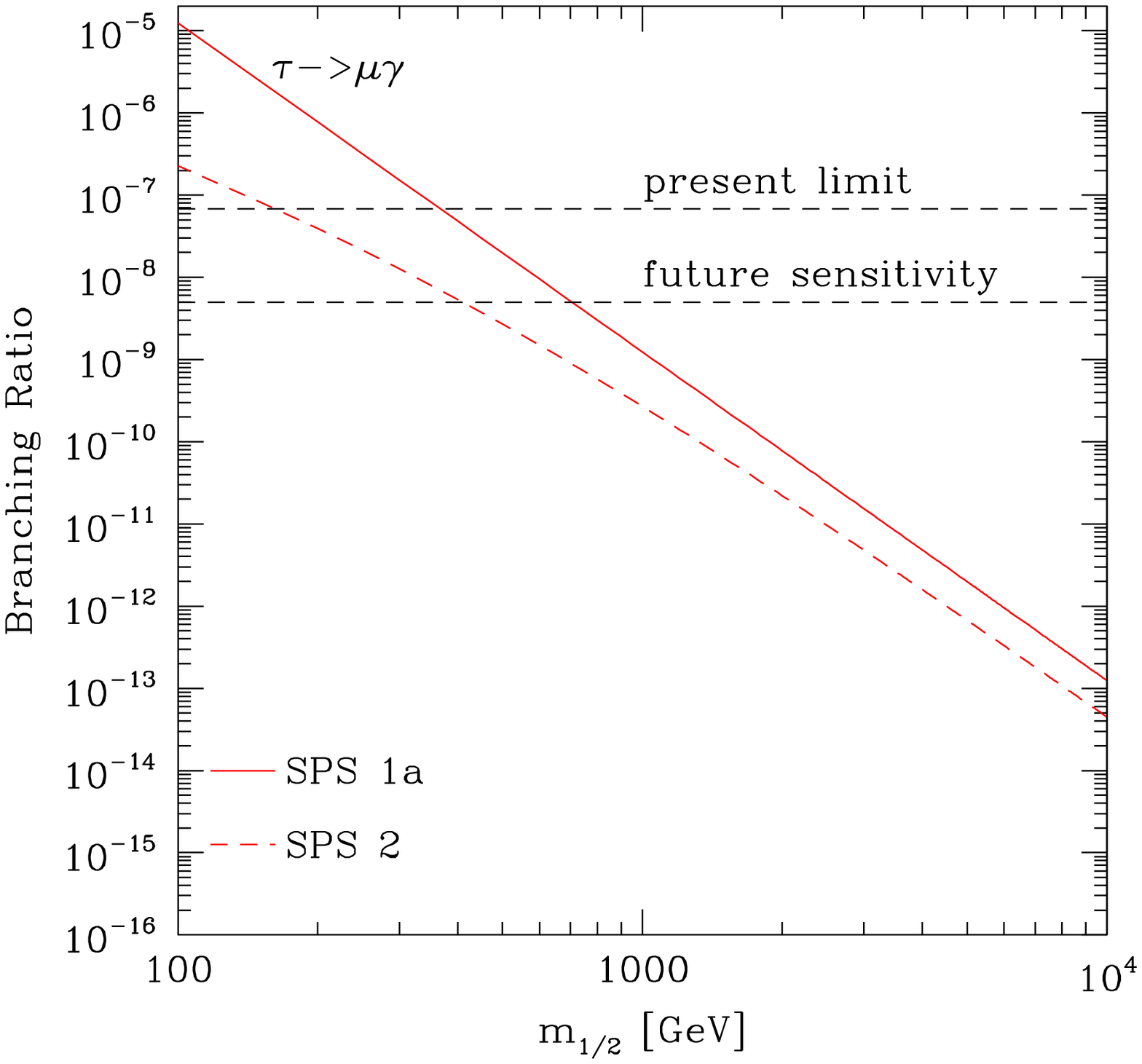,width=5.6cm,height=8cm}
\caption{\label{fig:BR2s}Second realization of QLC: 
the branching ratios for $\mu \ra e \gamma$, 
$\tau \ra e \gamma$ and $\tau \ra \mu \gamma$ against 
the SUSY parameter $m_{1/2}$ for the SPS slopes 1a and 2,  
see Table \ref{tab:SPS}. 
We have chosen for the neutrino mass (normal ordering) 0.02 eV. 
Indicated are also the present and future sensitivities.}
\end{figure}

With the help of Eqs.~(\ref{eq_ijg}, \ref{eq:mdt2}) we can evaluate 
the branching ratios for LFV processes, 
ignoring for the moment the logarithmic dependence on the 
heavy neutrino masses. 
The decay $\mu \ra e  \gamma$ is found to be governed by 
\be \label{eq:mueg2}
\left| (\tilde{m}_D^\dagger  \, \tilde{m}_D )_{21}
\right|^2 \simeq \frac 14 \, A^2 \, m_t^4 \, \lambda^4 
+ {\cal O}(\lambda^{5})~.
\ee
Comparing with Eq.~(\ref{eq:mueg}) we see that in the second realization 
the branching ratio is larger than in the first realization 
by 6 inverse powers of $\lambda$, or $\lambda^{-6} \simeq 8820$, almost 
4 orders of magnitude.  
For the double ratios of the branching ratios we obtain 
\be \label{eq:double1}
R (21/31) \simeq 1 - 2 \sqrt{2} \,  A \, \cos(\sigma - \tau) \, \lambda^2 + 
{\cal O}(\lambda^{3})~~,~~
R (21/32) \simeq A^2 \, \lambda^4 
+ {\cal O}(\lambda^{5})~.
\ee
There is a small dependence on the phase combination ($\sigma - \tau$), 
which also governs leptonic $CP$ violation in oscillation experiments 
and the magnitude of the atmospheric neutrino mixing angle. 
The branching ratios behave according to 
\be \label{eq:behave1}
\text{BR}(\mu \ra e \gamma):\text{BR}(\tau \ra e \gamma):
\text{BR}(\tau \ra \mu \gamma) \simeq A^2 \, \lambda^4 : A^2 \, \lambda^4 : 1~.
\ee
In Fig.\ \ref{fig:BR2NH} we show the branching ratios for 
$\mu \ra e \gamma$, $\tau \ra e \gamma$ and $\tau \ra \mu \gamma$ 
as a function of the smallest neutrino mass for a normal mass 
ordering, choosing the SPS points 1a, 2 and 4. The small dependence on 
the heavy neutrino masses is taken into account and plots for the inverted 
ordering look very similar. Note that from Fig.~\ref{fig:BR2NH} 
it follows that the dependence on the neutrino masses is very small. 
The relative magnitude of the branching ratios, as 
estimated in Eq.~(\ref{eq:double1}), holds true to a very high accuracy.
However, we immediately see that the prediction for $\mu \ra e \gamma$ is 
at least one order of magnitude above the current limit. 
To obey the experimental limit on BR($\mu \ra e \gamma$), the SUSY masses 
should be in the several TeV range. This is illustrated in 
Fig.~\ref{fig:BR2s}, where we have plotted the branching ratios 
as a function of the SUSY parameter $m_{1/2}$ for 
the SPS slopes 1a and 2 from Table \ref{tab:SPS}. 
We took the normal ordering of neutrino masses with 
a smallest mass $m_1 = 0.02$ eV. 
Once we have adjusted the SUSY parameters to have BR($\mu \ra e \gamma$) 
below its current limit, the other decays $\tau \ra e \gamma$ 
and $\tau \ra \mu \gamma$ are too rare to be observed with 
presently planned experiments. 

\section{\label{sec:concl}Conclusions and Summary}
We have considered the phenomenology of two predictive see-saw 
scenarios leading approximately 
to Quark-Lepton Complementarity. Both have in common that 
bimaximal mixing is corrected by the CKM matrix. 
We have studied the complete low energy phenomenology, including the neutrino 
oscillation parameters, where we have taken into account all 
possible phases, and neutrinoless double beta decay. 
Moreover, lepton flavor violating charged lepton decays have been 
studied and all results have been compared to presently available 
and expected future data. 
Finally, we have commented on the possibility of 
leptogenesis\footnote{As indicated at the 
beginning of Section \ref{sec:LFV0}, the decays 
$\mu \ra e \gamma$, $\tau \ra e \gamma$ and 
$\tau \ra \mu  \gamma$ are very strongly suppressed and can not be observed 
if supersymmetry is not realized by Nature. Hence, judging the 
validity of a given see-saw scenario based on its predictions for 
those decays can in this case not be done. 
Note that the predictions for leptogenesis do not depend on 
the presence of supersymmetry.} .\\ 

In terms of the elements of the PMNS matrix $U$ 
and the CKM matrix $V$, the QLC condition 
can be written as $|U_{e2}| + |V_{ud}| = 1/\sqrt{2}$. 
This defines the solar neutrino mixing parameter $\sin^2 \theta_{12}$ 
to be $\sin^2 (\frac{\pi}{4} - \lambda)$. Taking the best-fit, 
as well as the 1, 2 and 3$\sigma$ values of $\lambda$ from 
Eq.~(\ref{eq:ckm}) we obtain
\be \label{eq:QLCexact}
\sin^2 \theta_{12} = 0.2805 \pm (0.0009, \, 0.0018, \, 0.0027)~.
\ee 
A second QLC relation has also been suggested, namely 
$\theta_{23} + A \, \lambda^2 = \pi/4$, which is the 
analogue of Eq.~(\ref{eq:QLC}) for the (23)-sector. 
This can also be written as $|U_{\mu 3}| + |V_{cb}| = 1/\sqrt{2}$ and 
its precise prediction is 
\be \label{eq:QLCexact2}
\sin^2 \theta_{23} = 0.4583^{+ 0.0011, \, 0.0022, \, 0.0032}
_{-0.0011, \, 0.0022, \, 0.0034}~.
\ee
We remark that in our scenario with all possible $CP$ phases 
the above two relations correspond to at least one phase being zero.\\

\begin{figure}[htb]
\epsfig{file=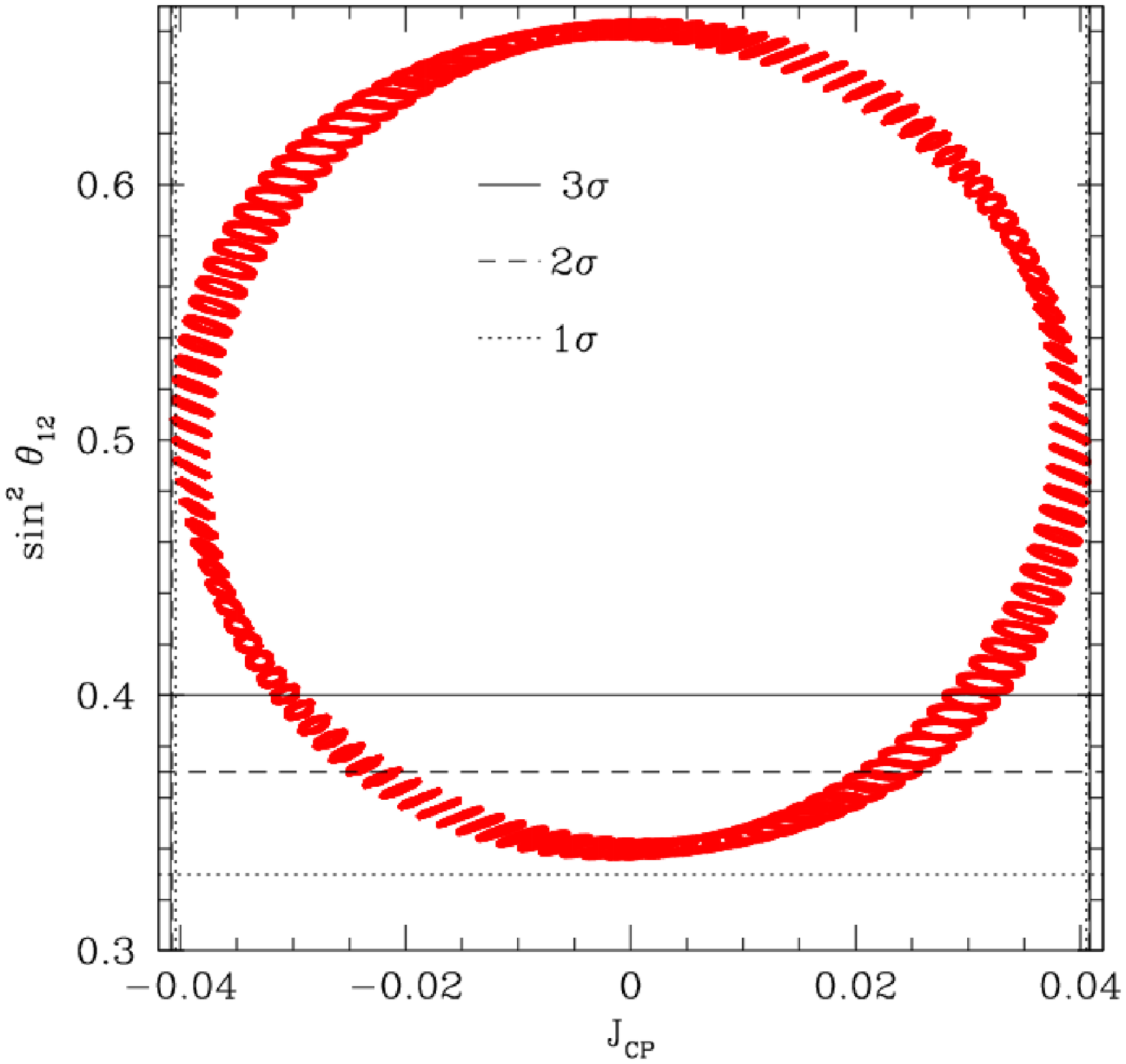,width=8.5cm,height=9cm}
\epsfig{file=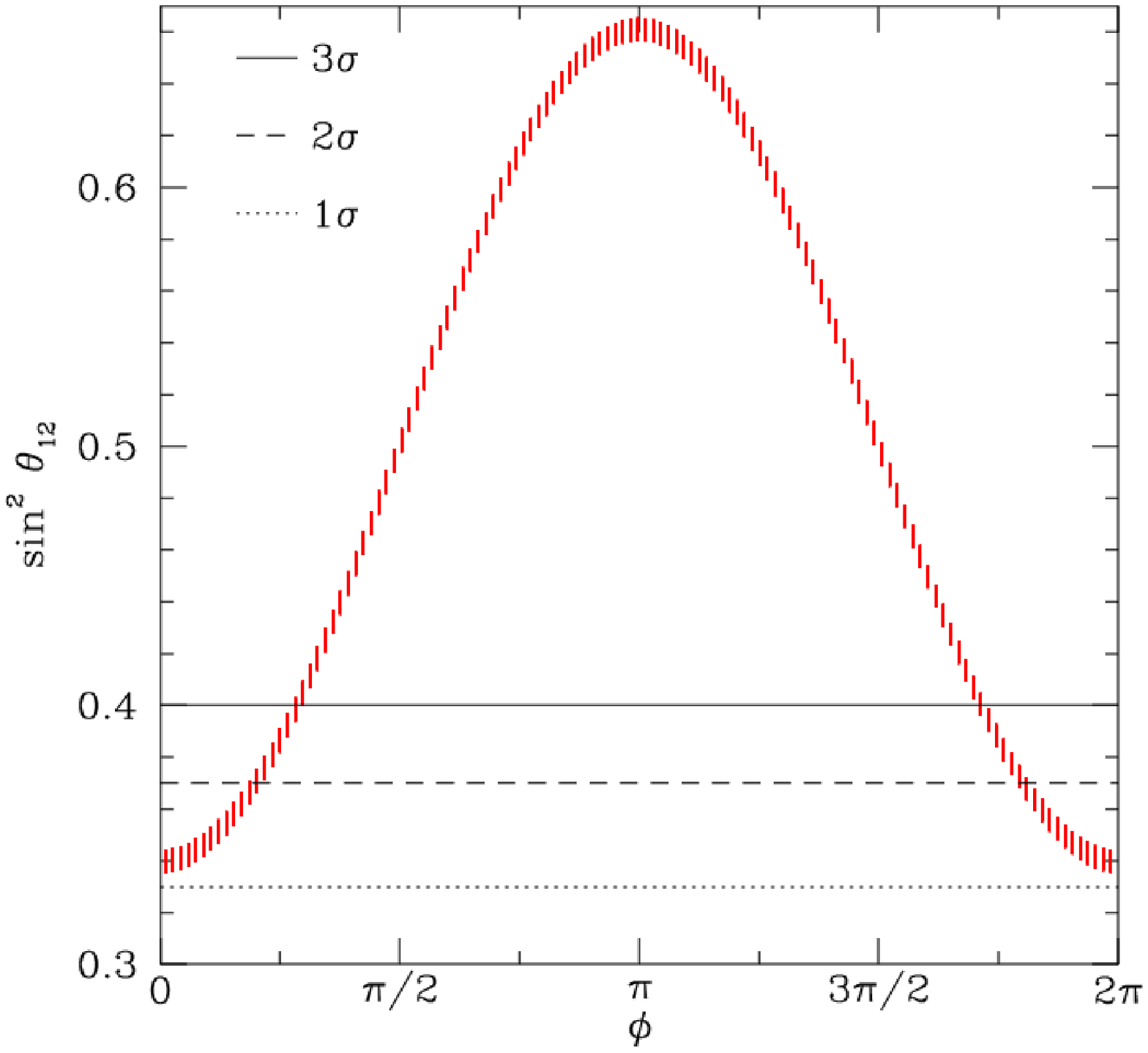,width=8.5cm,height=9cm}
\epsfig{file=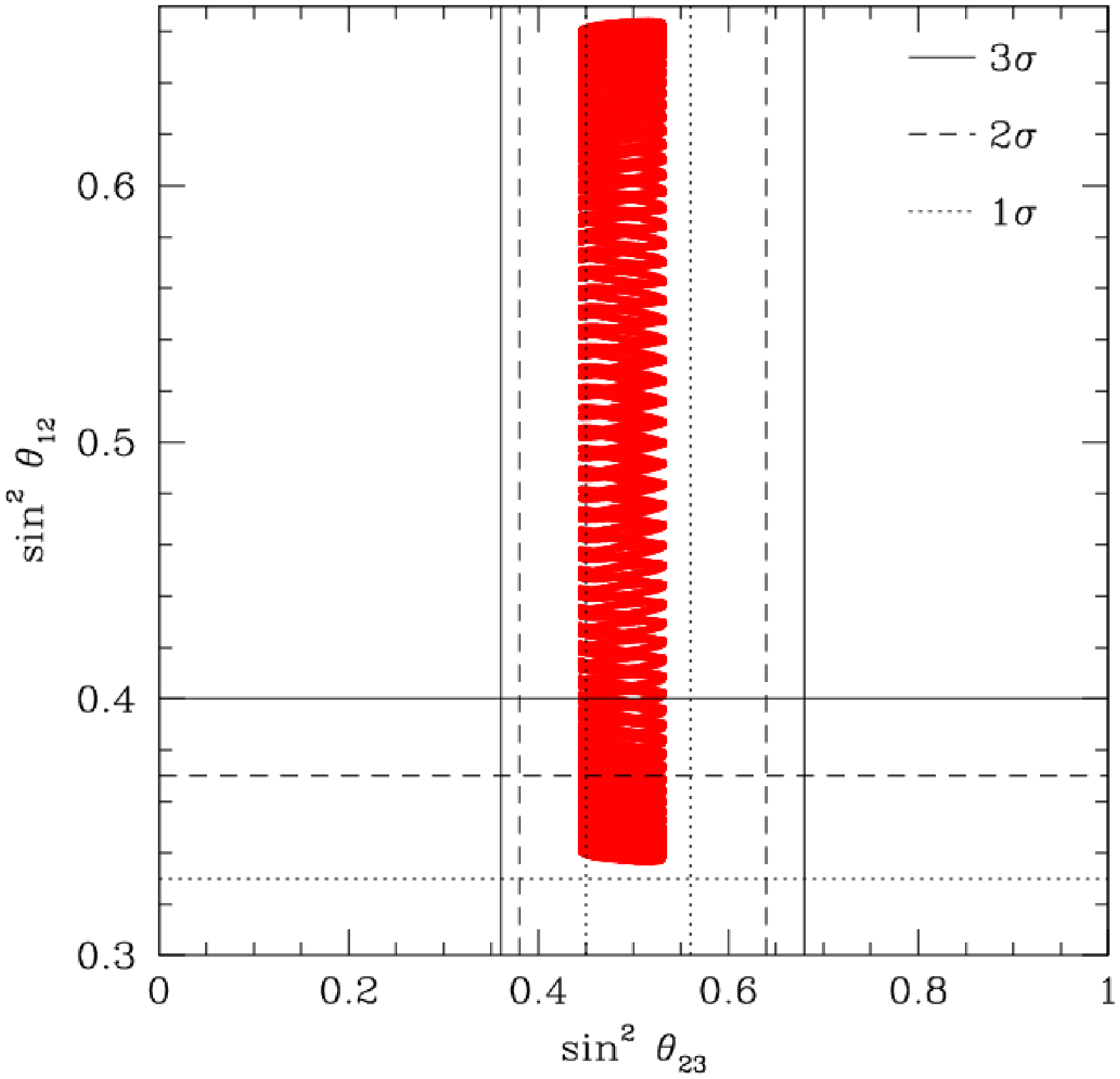,width=8.5cm,height=9cm}
\epsfig{file=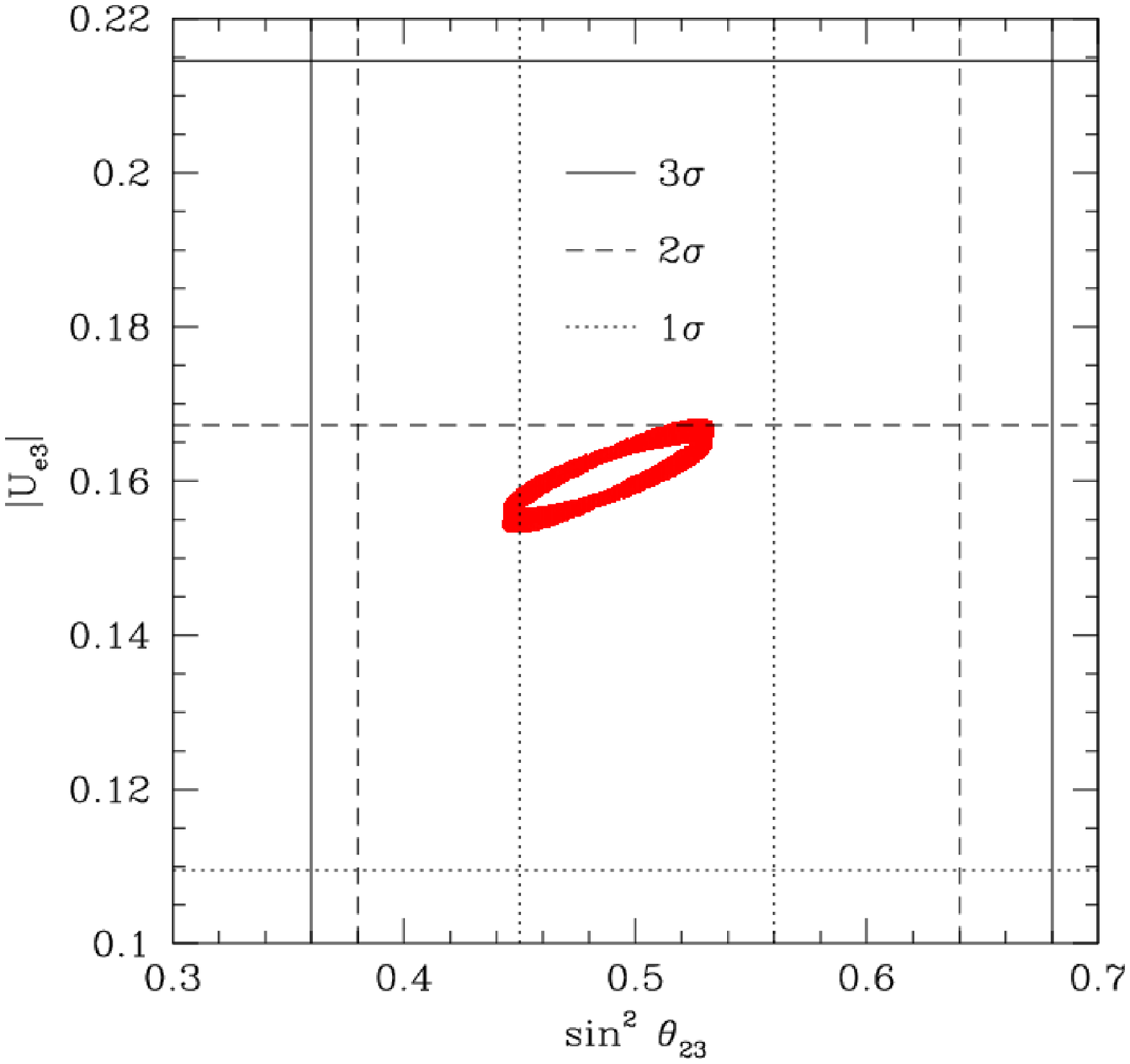,width=8.5cm,height=9cm}
\caption{\label{fig:the_one_ring}First realization of QLC: 
neutrino observables resulting from 
Eq.~(\ref{eq:QLCobs}) for the 3$\sigma$ ranges of the CKM 
parameters. We also indicated the present 
1, 2 and 3$\sigma$ ranges of the oscillation 
parameters.}
\end{figure}

The first scenario has bimaximal mixing from the neutrino sector and the 
matrix diagonalizing the charged leptons is the CKM matrix. 
The main results are:  
\begin{itemize}
\item  
solar neutrino mixing is predicted close to its 1$\sigma$ bound 
and $|U_{e3}|$ even close to its 2$\sigma$ bound, 
see Fig.~\ref{fig:the_one_ring}. 
The phase governing the magnitude of 
$\theta_{12}$ is the $CP$ phase of neutrino oscillations and is implied 
to be small; 
\item $|U_{e3}|$ is roughly 0.16, i.e., it should be observed soon; 
\item the lowest value of $\sin^2 \theta_{12}$ 
(corresponding to $CP$ conservation) is roughly 0.33, which differs 
by about 15\% from Eq.~(\ref{eq:QLCexact}). 
For $\sin^2 \theta_{23}$ the lowest value is 0.44, in moderate 
agreement with Eq.~(\ref{eq:QLCexact2}); 
\item the decay $\mu \ra e \gamma$ can be observable for not too 
small neutrino masses, unless the SUSY masses approach the TeV scale. 
BR($\tau \ra e \gamma$) is predicted to be very small, and 
observation of $\tau \ra \mu \gamma$ requires rather large neutrino masses, 
small SUSY masses, or large $\tan \beta$. 
The relative magnitude of the branching ratios 
is $\text{BR}(\mu \ra e \gamma):\text{BR}(\tau \ra e \gamma):
\text{BR}(\tau \ra \mu \gamma) \simeq \lambda^6 :  
\lambda^2 : 1$; 
\item successful resonant leptogenesis depends on the low energy Majorana 
phases but is fine-tuned. 
One possibility occurs  
if $\sigma \simeq \pi/2$, leading to two quasi-degenerate heavy neutrinos 
masses. It also leads to quasi-degenerate light neutrinos with 
mass around 0.5 eV and to sizable 
cancellations in \onbb$\!\!$, corresponding to $\meff \simeq 0.16$ eV.  
Leptogenesis via the decay of the second heaviest neutrino 
typically fails, even with the inclusion of flavor effects.  

\end{itemize}

\begin{figure}[htb]
\epsfig{file=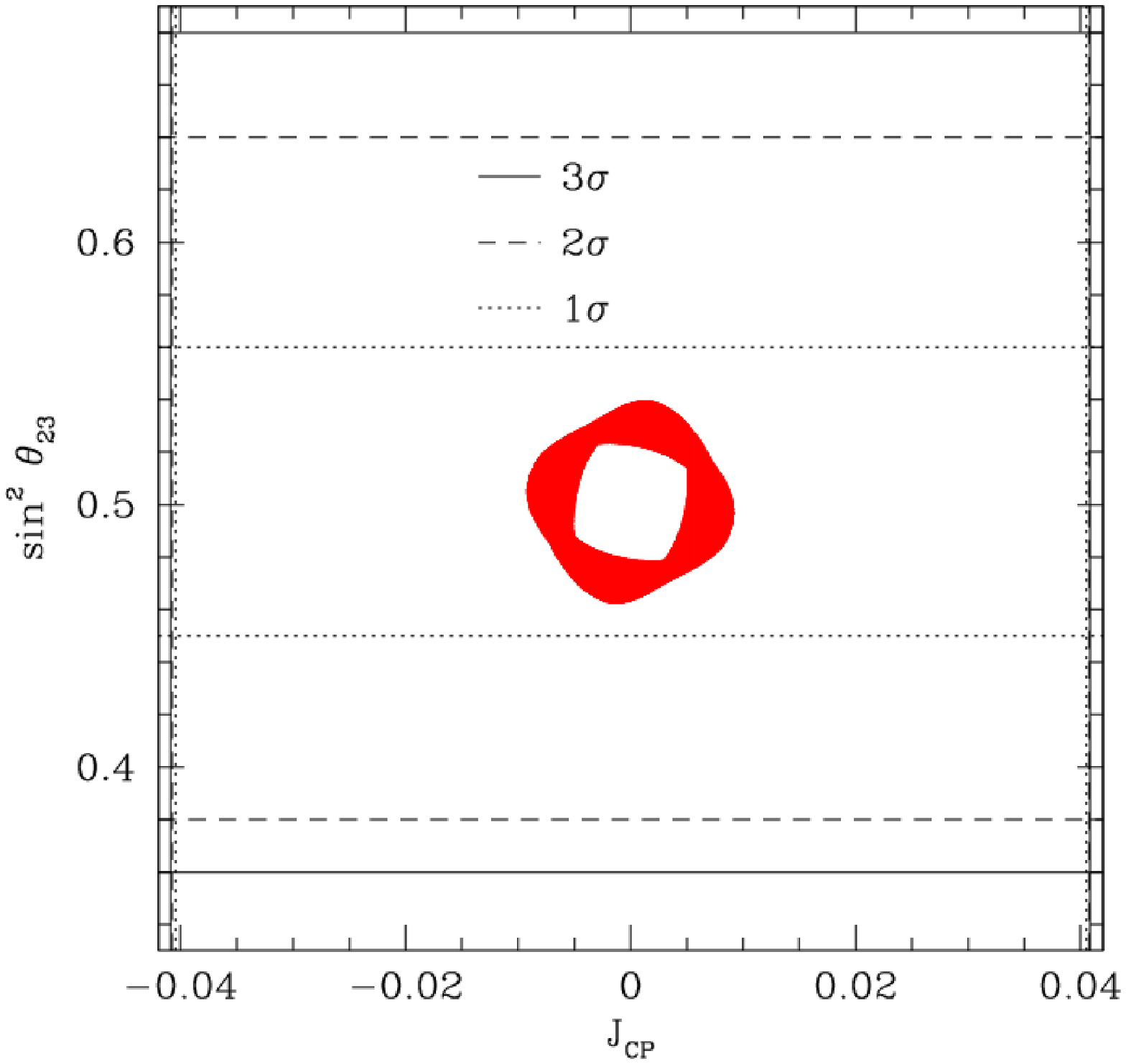,width=8.5cm,height=9cm}
\epsfig{file=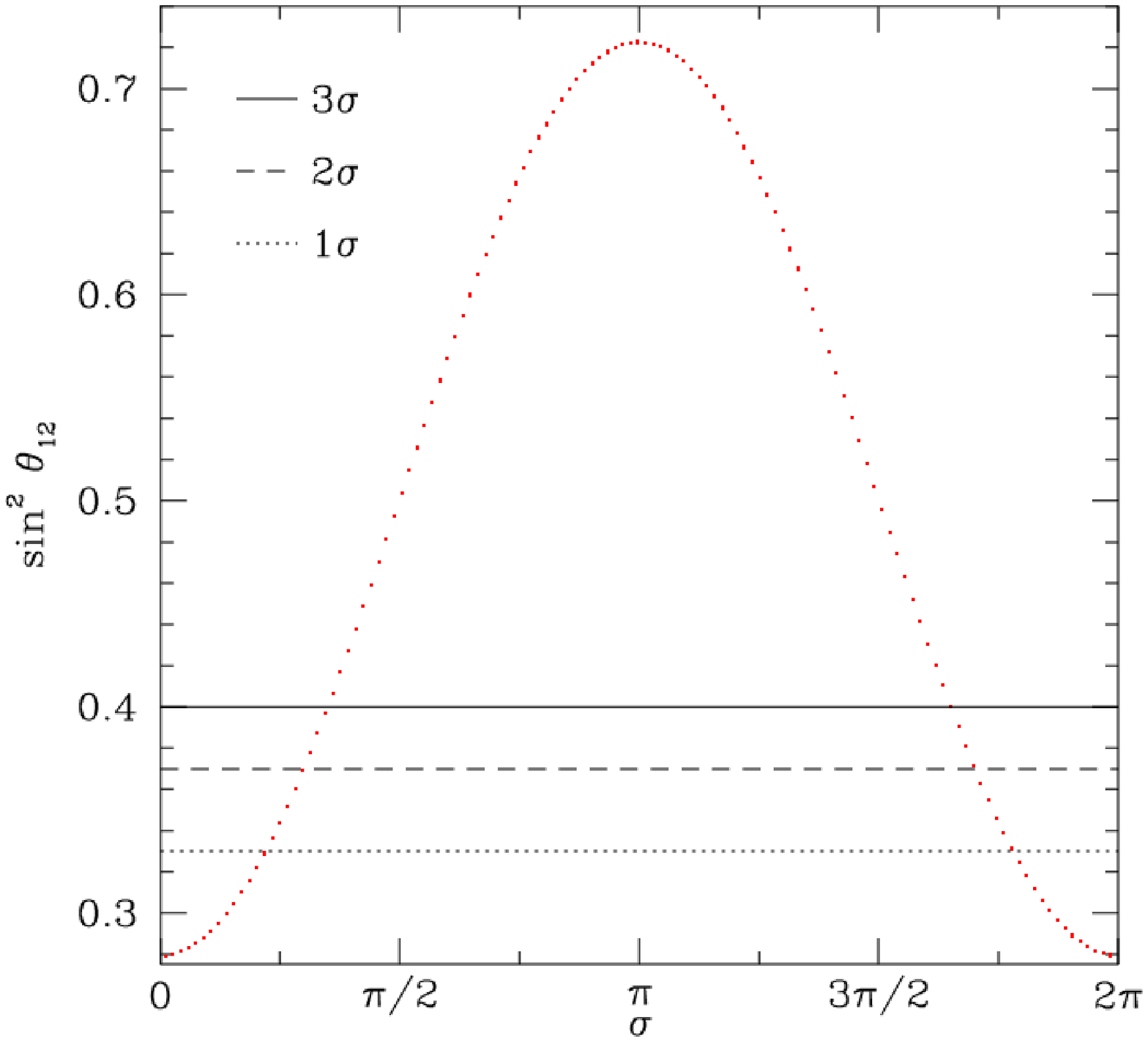,width=8.5cm,height=9cm}
\epsfig{file=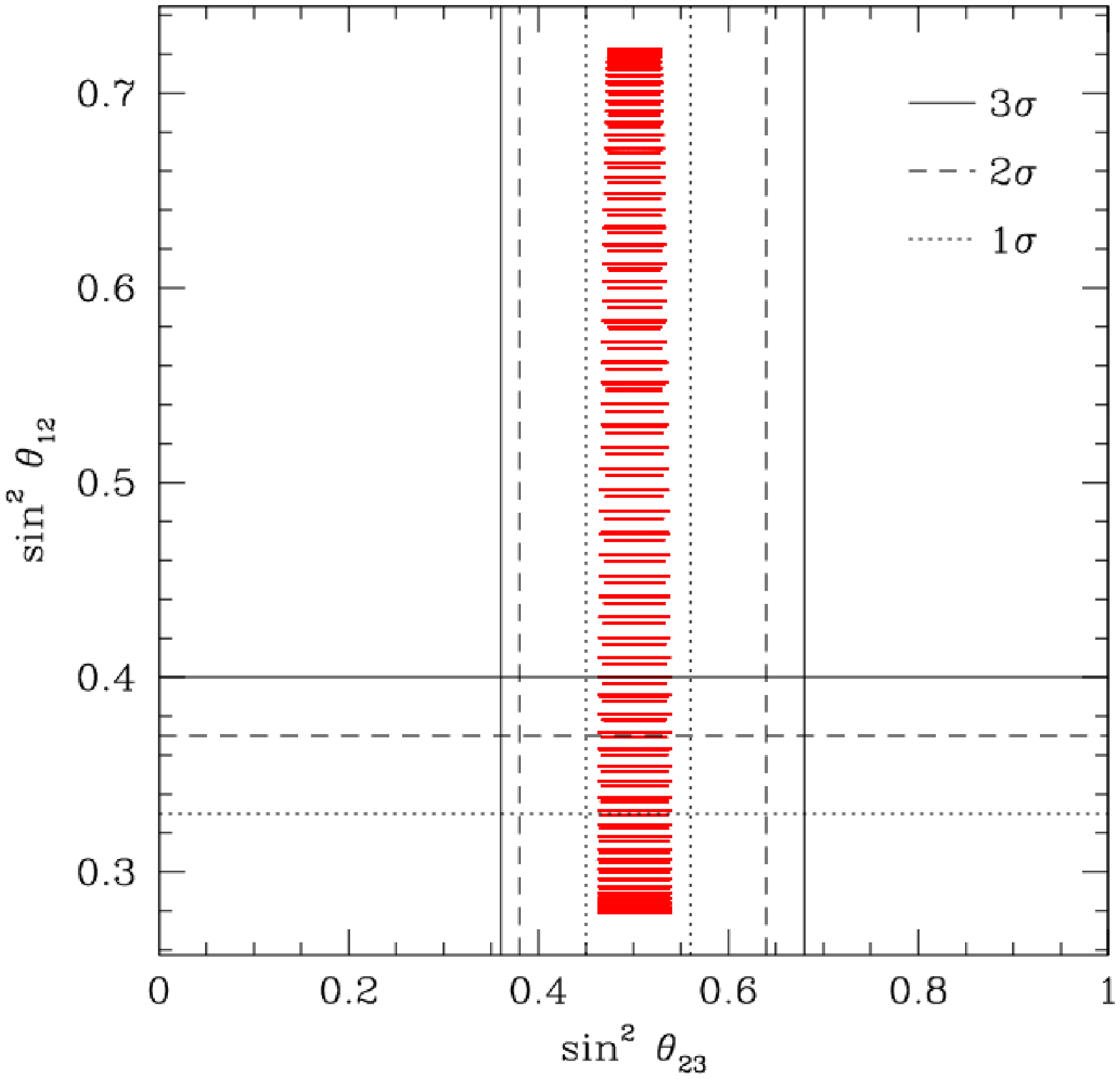,width=8.5cm,height=9cm}
\epsfig{file=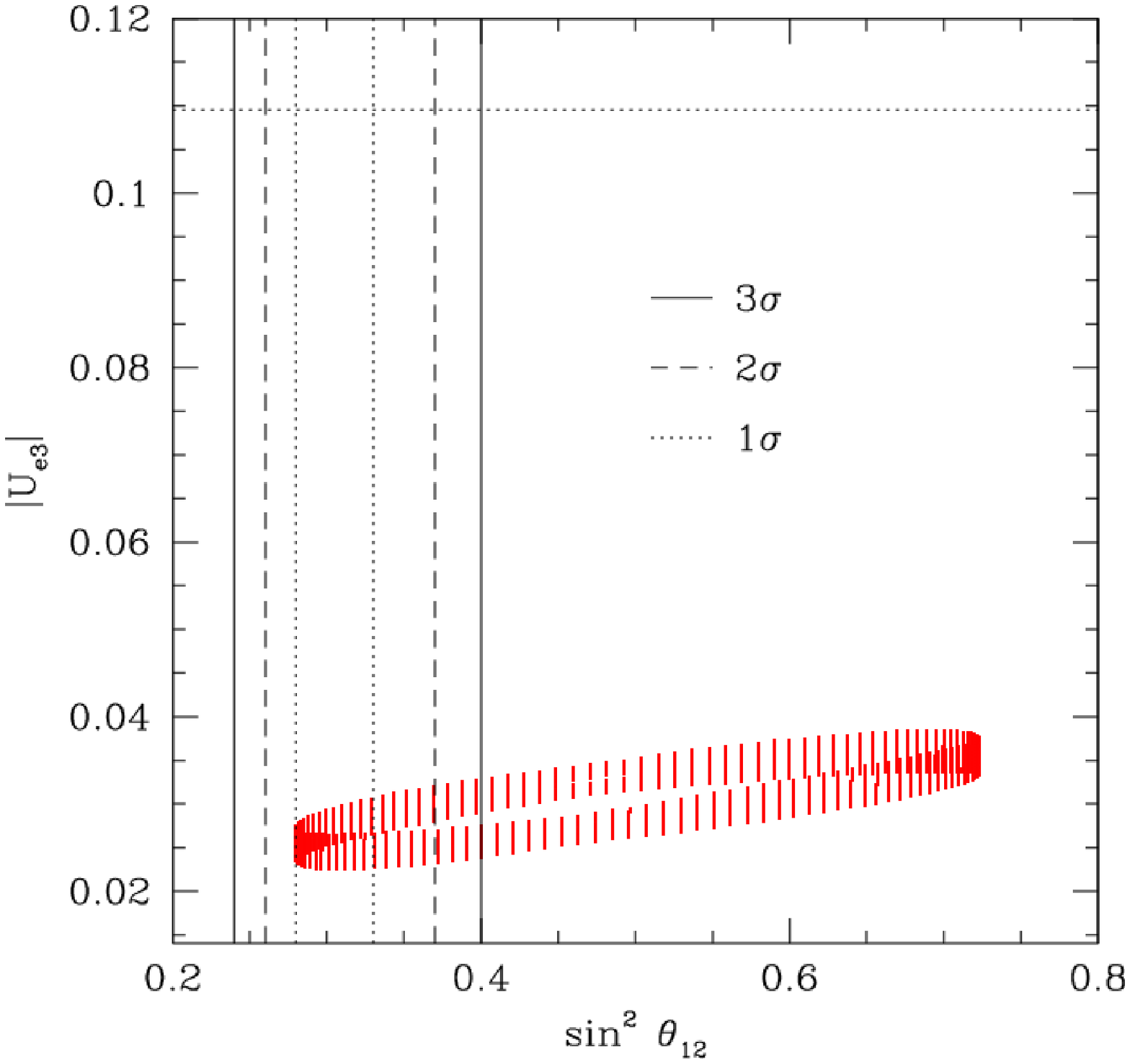,width=8.5cm,height=9cm}
\caption{\label{fig:the_one_ring2}Second realization of QLC: 
neutrino observables resulting from 
Eq.~(\ref{eq:QLCobs2}) for the 3$\sigma$ ranges of the CKM 
parameters. We also indicated the current 
1, 2 and 3$\sigma$ ranges of the oscillation 
parameters.}
\end{figure}

The second scenario has bimaximal mixing from the charged lepton sector 
and the matrix diagonalizing the neutrinos is the CKM matrix.
The main results are:  
\begin{itemize}
\item the neutrino oscillation parameters are perfectly compatible 
with all data, see Fig.~\ref{fig:the_one_ring2}. 
The phase governing the magnitude of $\theta_{23}$ is the $CP$ 
phase of neutrino oscillations; 
\item $|U_{e3}|$ is roughly 0.03, which is a rather small 
value setting a challenge for future experiments; 
\item the lowest value of $\sin^2 \theta_{12}$ 
(corresponding to $CP$ conservation) is roughly 0.28, in perfect 
agreement with Eq.~(\ref{eq:QLCexact}). 
For $\sin^2 \theta_{23}$ the lowest value is 0.46 (but $\theta_{23}$ 
can be maximal), in perfect agreement with Eq.~(\ref{eq:QLCexact2}). 
If $\sin^2 \theta_{23} = \frac 12$ 
then maximal $CP$ violation is implied;  
\item The branching ratio of $\mu \ra e \gamma$ is larger than in the first 
scenario by six inverse powers of $\lambda$ and therefore typically 
too large unless the SUSY masses are of several TeV scale. 
If they are so heavy that $\mu \ra e \gamma$ is below its current limit, 
$\tau \ra e \gamma$ and $\tau \ra \mu \gamma$ are too small to be observed. 
The relative magnitude of the branching ratios 
is $\text{BR}(\mu \ra e \gamma):\text{BR}(\tau \ra e \gamma):
\text{BR}(\tau \ra \mu \gamma) \simeq A^2 \, \lambda^4 : A^2 \, 
\lambda^4 : 1$;  
\item there can be no leptogenesis.
\end{itemize}
We conclude that both scenarios predict interesting and easily testable 
phenomenology. 
However, the first scenario is in slight disagreement with 
oscillation data and allows leptogenesis only for fine-tuned 
parameter values. In the second 
scenario the predictions for LFV decays are in contradiction to 
experimental results unless the SUSY parameters are very large. 
Moreover no leptogenesis is possible in this case.

\vspace{0.5cm}
\begin{center}
{\bf Acknowledgments}
\end{center}
We are grateful to St\'ephane Lavignac for helpful discussions and 
thank J.~Klinsmann for invaluable encouragement. 
This work was supported by 
the ``Deutsche Forschungsgemeinschaft'' in the 
``Sonderforschungsbereich 375 f\"ur Astroteilchenphysik'' 
and under project number RO--2516/3--1 (W.R.).

\end{document}